\documentclass[%
 reprint,
superscriptaddress,
 amsmath,amssymb,
 aps,
prb,
 citeautoscript,
 10pt
]{revtex4-2}

\usepackage{extsizes}
\usepackage[version=3]{mhchem}
\usepackage[left=1.5cm, right=1.5cm, top=1.785cm, bottom=2.0cm]{geometry}
\usepackage{balance}
\usepackage{mathptmx}
\usepackage{graphicx} 
\usepackage{lastpage}
\usepackage[format=plain,justification=justified,singlelinecheck=false,font={stretch=1.125,small,sf},labelfont=bf,labelsep=space]{caption}
\usepackage{float}
\usepackage{fancyhdr}
\usepackage{fnpos}
\usepackage[english]{babel}
\addto{\captionsenglish}{%
  \renewcommand{\refname}{Notes and references}
}
\usepackage{array}
\usepackage{droidsans}
\usepackage{charter}
\usepackage[T1]{fontenc}
\usepackage[usenames,dvipsnames]{xcolor}
\usepackage{setspace}
\usepackage[compact]{titlesec}
\usepackage{hyperref}
\usepackage{siunitx}
\usepackage{nicefrac}
\usepackage{upgreek}
\usepackage{multirow}
\usepackage{braket}

\usepackage{epstopdf}

\DeclareSIUnit{\atom}{atom}

\DeclareSIUnit\angstrom{\text {Å}}

\DeclareMathAlphabet{\mathcal}{OMS}{cmsy}{m}{n}

\begin{document}

\title{Multiplet structure of chromium(III) dopants in wide band gap materials}

\author{Ilya Popov}
\affiliation{School of Chemistry, University of Nottingham, Nottingham, NG7 2RD, United Kingdom}

\author{Petros-Panagis Filippatos}
\affiliation{School of Chemistry, University of Nottingham, Nottingham, NG7 2RD, United Kingdom}

\author{Shayantan Chaudhuri}
\affiliation{School of Chemistry, University of Nottingham, Nottingham, NG7 2RD, United Kingdom}

\author{Andrei L. Tchougr\'{e}eff}
\affiliation{Frumkin Institute of Physical Chemistry and Electrochemistry of the Russian Academy of Sciences, Moscow, 119071, Russia}

\author{Katherine Inzani}
\affiliation{School of Chemistry, University of Nottingham, Nottingham, NG7 2RD, United Kingdom}

\author{Elena Besley}
\affiliation{School of Chemistry, University of Nottingham, Nottingham, NG7 2RD, United Kingdom}
\email{Elena.Besley@nottingham.ac.uk}

\begin{abstract}
Transition metal doping is commonly used for altering the properties of solid-state materials to suit applications in science and technology. Partially filled $d$-shells of transition metal atoms lead to electronic states with diverse spatial and spin symmetries. Chromium(III) cations have shown great potential for designing laser materials and, more recently, for developing spin qubits in quantum applications. They also represent an intriguing class of chemical systems with strongly correlated multi-reference excited states, due to the $d^3$ electron configuration. These states are difficult to describe accurately using single-reference quantum chemical methods such as density functional theory (DFT), the most commonly used method to study the electronic structures of solid-state systems. Recently, the periodic effective Hamiltonian of crystal field (pEHCF) method has been shown to overcome some limitations arising in the calculations of excited $d$-states. In this work, we assess the suitability of DFT and pEHCF to calculate the electronic structure and $d$–$d$ excitations of chromium(III) dopants in wide band gap host materials. The results will aid computational development of novel transition metal-doped materials and provide a deeper understanding of the complex nature of transition metal dopants in solids.
\end{abstract}

\maketitle

\section{Introduction}

Solid-state materials containing transition metal dopants are widely used in many areas of science and engineering, including optoelectronics~\cite{Suo_2020, Egbo_2021}, optics~\cite{Greeshma_2025, Dhale_2025}, laser technologies~\cite{McKinnie_1996,Carrig_2005,Alpaslan_Yayli_2021}, photovoltaics and photoelectrochemistry~\cite{Wieghold_2021,MARUSKA1979}, spintronics~\cite{Dierolf_2016_book}, semiconductors~\cite{Dierolf_2016_book,Murai_2021,Lopez_2025}, and quantum computing~\cite{sewani2020spin,Shang_2022}. The distinct properties of transition metal atoms primarily stem from the partially filled $d$-shells, which lead to a rich multiplet structure of the electronic spectrum, including states with different spatial and spin symmetries and respective degrees of degeneracy. The structure of $d$-multiplets can be fine-tuned by targeting doping sites with particular symmetries and modifying the host material to affect dopant--host interactions. Chromium(III) (\ce{Cr^3+}) is an attractive dopant choice as its $d^3$ electronic configuration and corresponding multiplet structure make it a useful candidate for laser materials~\cite{MAIMAN_1960,Sennaroglu_2021} and spin qubits~\cite{sewani2020spin}. For example, ~\citet{sewani2020spin} demonstrated optical initialization and read-out, along with long relaxation times, for the $S = \nicefrac{3}{2}$ spin population associated with \ce{Cr^3+} dopants in \ce{Al2O3}, where $S$ is the spin quantum number. Continued investigations of \ce{Cr^3+} and other transition metal dopants in different host materials will prove fruitful for quantum technologies by providing a rich set of systems for extended applications. 

Dopants with open $d$-shells exhibit electronic correlation effects~\cite{L_wdin_1955} of both static and dynamical nature, which occur due to the strong electron--electron interactions characteristic of \textit{d}-electrons. Static correlations appear when the ground state (or an excited state of interest) exhibits a multi-reference character, i.e. it cannot be represented quantitatively by a single Slater determinant (see ~\citet{Benavides_Riveros_2017} and references therein for further details). Such correlation effects are present in systems with electron (quasi-)degeneracy, which are typical for the $d$-states of transition metals in highly symmetrical crystal fields~\cite{Tanabe_1954}, either ideal or slightly distorted. These strongly-correlated materials pose a significant challenge to computational research~\cite{Lischner_2012,Pavarini_2021}  as the electronic structure of $d$-multiplets cannot be adequately captured by standard density functional theory~\cite{Hohenberg-Kohn, Kohn-Sham} (DFT) methods, which are commonly employed to investigate the electronic structures of solid-state systems. Limitations of DFT are most pronounced when calculating the $d$-$d$ excitations and band gaps of strongly-correlated systems. Advanced methods based on hybrid density functional approximations can yield better results than standard semi-local functionals~\cite{CzelejACSNano24}, however these solutions are not universal and have limitations on the types of excitations that can be accurately predicted~\cite{Yang_2023}.

This highlights the pressing requirement to search beyond DFT for more reliable methods for electronic structure calculations. One of the most direct ways for addressing strong correlations involves the use of post-Hartree-Fock approaches, such as complete active space self-consistent field (CASSCF)~\cite{OlsenIJQC11}, second-order M{\o}ller–Plesset perturbation theory~\cite{HEADGORDON1988503} and coupled cluster methods~\cite{Jeziorski_1981}. While examples of applying post-Hartree-Fock treatment to solid-state materials exist in the literature~\cite{Booth_2012,Gruber_2018,Neufeld_2022}, the range of applications is limited due to a strong scaling with system size. CASSCF has occasionally been used to calculate the energies of local $d$-$d$ excitations in small (finite) clusters that represent the first coordination sphere of a transition metal atom in a crystalline system~\cite{Shang_2022}. This approach is, however, associated with significant underestimation of the delocalization effects taking place in extended systems. An alternative to post-Hartree-Fock approaches is to combine multiple electronic structure methods within a hybrid embedding approach, whereby the electronic structure of localized $d$-shells is treated using a correlated method and the host is described within a weakly correlated (e.g. one-electron self-consistent field) treatment. Such hybrid approach allows one to reduce computational costs while providing a multi-reference description of strongly correlated \textit{d}-electrons. One of the examples of hybrid electronic structure methods used in solid state theory is dynamical mean-field theory~\cite{GeorgesRMP96}.

In our previous work~\cite{Popov_2023_1} we extended a hybrid embedding method, called the effective Hamiltonian of crystal field (EHCF),~\cite{Soudackov_1992,Tchougr_eff_2015} to periodic systems (pEHCF). pEHCF has been shown to be successful in describing the $d$-multiplet structure of various solid materials including oxides~\cite{Popov_2023_1}, carbodiimides and metal-organic frameworks~\cite{Popov_2023_mof,Popov_2025}.
In this work, we evaluate and compare the suitability of DFT and pEHCF to investigate the electronic structure and $d$–$d$ excitations of \ce{Cr^3+} dopants in three wide band gap host materials: corundum ($\upalpha$-\ce{Al2O3}), aluminium oxonitridoborate (\ce{AlB4O6N})~\cite{Widmann_2024} and chrysoberyl (\ce{BeAl2O4}).

$\upalpha$-\ce{Al2O3} with \ce{Cr^3+} dopants is a well-known laser material~\cite{MAIMAN_1960,Ratzker_2021} with excellent optical properties and distinctive fluorescence which has been also explored for quantum applications~\cite{sewani2020spin}. \ce{AlB4O6N} recently synthesized by ~\citet{Widmann_2024} possesses interesting fluorescence and luminescence properties as well as high thermal stability. Finally, \ce{BeAl2O4} has been widely utilized in solid-state laser technologies due to its exceptional emission properties in the 700--\SI{800}{\nm} range~\cite{scalvi2005thermal}. All three host materials show characteristics that are promising for quantum technologies which will benefit from deeper understanding of the complex electronic structure of transition metal dopants in solids provided in this study.

\section{Methodology}

\subsection{Periodic Effective Hamiltonian of Crystal Field}

The periodic effective Hamiltonian of crystal field (pEHCF) method splits the electronic system into two subsystems, where one contains only the local $d$-shell(s) of transition metal atoms and the other includes the crystalline environment embedding these shells~\cite{Popov_2023_1}. The idea of such division of the electronic system into subsystems was first proposed by ~\citet{Harrison1980}, who considered the electronic structure of transition metal oxides as delocalized $s,p$-bands augmented with local $d$-multiplets. Mathematically, this splitting is performed by separating a space of one-electron states into $d$- and $l$-subspaces spanned by local atomic $d$- and $s,p$-orbitals, respectively. Transition metal $s,p$-orbitals are included in the $l$-subsystems along with the orbitals of light elements. The many-electron wavefunction of the system is expressed using Equation~(\ref{eq:WF-total}): 


\begin{equation}\label{eq:WF-total}
\varPsi=\varPsi_{d}\left(n_{d}\right)\wedge\varPsi_{l}\left(N-n_{d}\right)
\end{equation}
where $\varPsi_{d}$ and $\varPsi_{l}$ are the many-electron wavefunctions built in the $d$- and $l$- subspaces, respectively, $n_{d}$ is the number of electrons in the $d$-subspace, $N$ is the total number of electrons in the system, and $\wedge$ stands for the antisymmetric product. The wavefunctions $\varPsi_{d}$ and $\varPsi_{l}$ are treated differently. In the case of an impurity ion, its strongly correlated $d$-shell is described by a full configuration interaction wavefunction:
\begin{equation}\label{eq:WF_d}
\varPsi_{d}\left(n_{d}\right)=\sum_{i}c_{i}\varPsi_{i}\left(n_{d}\right)
\end{equation}
where the summation goes over all configurations, $\varPsi_{i}$, with coefficients $c_i$, built on $d$-orbitals of the impurity ion. This accounts for correlations in the $d$-shell and accurately reproduces both the energy and spin multiplicity of the excited $d$-multiplets. The non-correlated $l$-subsystem is treated using the Hartree-Fock method with the single-determinant wavefunction in the basis of the Bloch states constructed from the $s,p$-atomic orbitals of the system~\cite{Popov_2023_1}. 

The wavefunction in Equation~(\ref{eq:WF-total}) assumes the number of electrons in both $d$- and $l$-subsystems to be fixed and therefore excludes charge transfer states. The presence of such states is taken into account in pEHCF by the L{\"o}wdin partitioning technique~\cite{L_wdin_1982}, which provides effective corrections to the Hamiltonian operators of the subsystems arising due to the electron hopping between them~\cite{Soudackov_1992, Popov_2023_1}. This results in the effective Hamiltonian for the $d$-subsystem, $\mathcal{H}_d^\mathrm{eff}$, having the following form:
\begin{eqnarray}\label{eq:Heff_d}
\mathcal{H}_{d}^\mathrm{eff}&=&\sum_{\mu\nu}\sum_{\sigma}\left(\mathcal{H}_{\mu\nu}^{d}+\mathcal{H}_{\mu\nu}^\mathrm{coul}+\mathcal{H}_{\mu\nu}^\mathrm{res}\right)d_{\mu\sigma}^{+}d_{\nu\sigma}+\notag \\
 &+& \frac{1}{2} \sum_{\mu\nu\lambda\eta}\sum_{\sigma\tau}(\mu\nu|\lambda\eta)d_{\mu\sigma}^{+}d_{\lambda\tau}^{+}d_{\eta\tau}d_{\nu\sigma}+\text{h.c.}
\end{eqnarray}
where $d_{\mu\sigma}^{+}$ and $d_{\mu\sigma}$ are the electron creation and annihilation operators, respectively, for the $\mu$-th $d$-orbital, $\sigma$ and $\tau$ correspond to the spin projection, and $\text{h.c.}$ is the Hermitian conjugate. The second term includes two-electron one-center integrals $(\mu\nu|\lambda\eta)$ over atomic $d$-orbitals and describes the two-electron Coulomb interactions within the $d$-shell. The one-electron part of $\mathcal{H}_d^\mathrm{eff}$ includes the bare Hamiltonian of the $d$-subsystem ($\mathcal{H}^{d}$), and both Coulomb ($\mathcal{H}^\mathrm{coul}$) and resonance ($\mathcal{H}^\mathrm{res}$) interactions of $d$-electrons with electrons and nuclei in the $l$-subsystem. Contributions from $\mathcal{H}^\mathrm{coul}$ and $\mathcal{H}^\mathrm{res}$ determine the `splitting parameter' of the $d$-orbitals, as referred to within the crystal field theory. In pEHCF, unlike in the crystal field theory, the main contribution to the splitting of $d$-orbitals is the resonance interactions~\cite{Soudackov_1992,Popov_2023_1} corresponding to the one-electron transfers between the $d$- and $l$-subsystems. Therefore, $\mathcal{H}^\mathrm{res}$ is the most important factor when analyzing variations in the splitting parameters during spin-crossover processes. The matrix elements of $\mathcal{H}^\mathrm{res}$ have the following form~\cite{Tch068,Popov_2023_1}:

\begin{equation}\label{eq:Resonance-term}
\mathcal{H}_{\mu\nu}^\mathrm{res}=\sum_{a,b}\beta_{\mu a}\beta_{\nu b}\bigg(\Re G_{ab}^{+}(-I_{d})+\Re G_{ab}^{-}(-A_{d})\bigg)
\end{equation}
where the summation is over atomic orbitals (AOs) $a$ and $b$ of the $l$-subsystem, $\beta_{\mu a}$ are resonance (hopping) integrals between the $\mu$-th $d$-orbital and $l$-AO $a$, and $I_d$ and $A_d$ are the ionization potential and electron affinity, respectively, of the $d$-subsystem. The orbital-projected Green's functions, $G_{ab}^{\pm}$, of the $l$-subsystem are expressed as:

\begin{equation}\label{eq:GF-plus}
G_{ab}^{+}\left(\varepsilon\right)=\lim_{\delta\rightarrow0^{+}}\sum_{n,\mathbf{k}}\left(1-f_{n\mathbf{k}}\right)\frac{\braket{a|n\mathbf{k}}\braket{n\mathbf{k}|b}}{\varepsilon-\varepsilon_{n\mathbf{k}}+i\delta}
\end{equation}

\begin{equation}\label{eq:GF-minus}
G_{ab}^{-}\left(\varepsilon\right)=\lim_{\delta\rightarrow0^{+}}\sum_{n,\mathbf{k}}f_{n\mathbf{k}}\frac{\braket{a|n\mathbf{k}}\braket{n\mathbf{k}|b}}{\varepsilon-\varepsilon_{n\mathbf{k}}+i\delta}
\end{equation}

In Equations~(\ref{eq:GF-plus}) and (\ref{eq:GF-minus}), $\mathbf{k}$ is a vector in the first Brillouin zone, $n$ enumerates bands of the $l$-subsystem, and $\varepsilon_{n\mathbf{k}}$ and $f_{n\mathbf{k}}$ are energies and occupation numbers, respectively, of the $l$-bands. Spin variables are omitted for clarity. As can be seen, the resonance term (and therefore the splitting of the $d$-orbitals) depends on three main factors: the geometry of the first coordination sphere through the resonance integrals between local atomic orbitals ($\beta_{\mu a}$), the occupations of local atomic orbitals forming the first coordination sphere, and the energy difference between $d$-states and the valence/conduction bands of the $l$-subsystem. 

Solving the linear Schr{\"o}dinger equation for the wavefunction of the $d$-system, as shown in Equation~(\ref{eq:WF_d}), with the effective Hamiltonian in Equation~(\ref{eq:Heff_d}) produces the whole spectrum of energies for the $d$-multiplets with all allowed spins and symmetries, among them the ground state. pEHCF electronic structure calculations were performed for geometries obtained by DFT with use of the r$^2$SCAN~\cite{furness2020accurate} meta-generalized gradient approximation as described below.

\subsection{Density Functional Theory Calculations}

Kohn-Sham DFT~\cite{Hohenberg-Kohn, Kohn-Sham} calculations were performed using version 6.4.1 of the Vienna Ab initio Simulation Package (VASP)~\cite{kresse1996efficient, kresse1999ultrasoft, kresse1996efficiency} software within the spin-polarized framework, with the Be ($s^{1.99}p^{0.01}$), B ($2s^{2}2p^{1}$), and O ($2s^{2}2p^{4}$), Al ($3s^{2}3p^{1}$), and Cr\_sv ($3s^{2}3p^{6}3d^{5}4s^{1}$) projector-augmented wave pseudopotentials~\cite{blochl1994projector, kresse1999ultrasoft}. The r$^2$SCAN~\cite{furness2020accurate} meta-generalized gradient approximation (meta-GGA) and the HSE06~\cite{heyd2003hybrid,paier2006screened} range-separated hybrid generalized gradient approximation were used to perform geometry optimization of the structures and their corresponding density of states. Converged plane-wave cut-off energies of \SI{700}{\electronvolt}, \SI{700}{\electronvolt}, and \SI{750}{\electronvolt} were used for the \ce{Al2O3}, \ce{BeAl2O4}, and \ce{AlB4O6N} hosts, respectively. Furthermore, a reciprocal space grid ($k$-grid) of size $4\times4\times2$  was used for the primitive unit cells. Convergence tests for the plane-wave cut-off energy and $k$-point mesh were performed until the total energies from single-point calculations were converged to within \SI{1}{\milli\electronvolt\per\atom}. All undoped structures were optimized until the residual forces on all ions were less than \SI{0.001}{\electronvolt\per\angstrom} while for the defected supercells the convergence criterion of forces on all ions was set to \SI{0.01}{\electronvolt\per\angstrom}. The energy convergence criterion for geometry optimizations was set to $10^{-6}$~\si{\electronvolt}. For the defect structures, $2\times2\times1$ expanded supercells of 120, 168, and 192 atoms were used for $\upalpha$-\ce{Al2O3}, \ce{BeAl2O4}, and \ce{AlB4O6N}, respectively, to reduce interactions between periodic images of the defects. Delta Self-Consistent Field ($\Delta$SCF) calculations were performed to determine the electronic excited states and the corresponding absorption energies~\cite{PhysRevLett.103.186404,RevModPhys.86.253}. This method is known to be highly effective for calculating excitation energies in quantum defects~\cite{wang2021spin}. The $\Delta$SCF occupation for the spin-up and spin-down components at each $k$-point was manually set to fix the electron configuration of the system. This provides the total energy of the excited structure without relaxing the geometry. We use this method to simulate the vertical excitations of the first and second excited states of  crystal hosts.

\section{Results and Discussion}

\begin{figure*}
\begin{centering}
\includegraphics[scale=0.6]{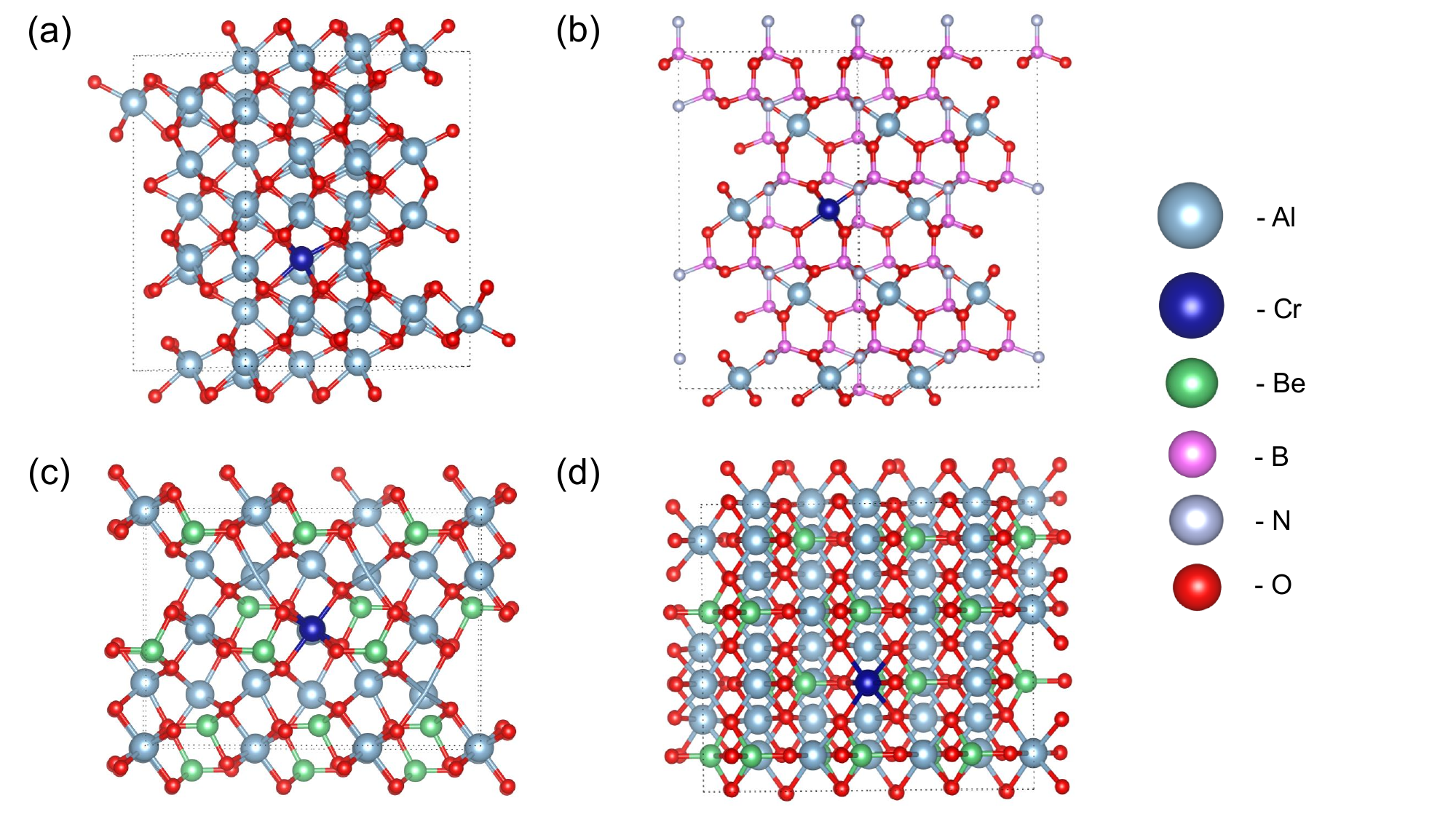}
\par\end{centering}
\caption{Crystalline structures of \ce{Cr^3+}-doped (a) $\upalpha$-Al$_2$O$_3$, (b) AlB$_4$O$_6$N, and BeAl$_2$O$_4$ at (c) $C_{s}$- and (d) $C_{i}$-symmetrized sites. Cell boundaries are shown with black dotted lines. All shown crystalline structures are available in Supplementary Information for both r$^{2}$SCAN and HSE06 functionals.\dag}
\label{fig:structures}
\end{figure*}

The crystal lattices of $\upalpha$-\ce{Al2O3}, \ce{AlB4O6N}, and \ce{BeAl2O4} correspond to trigonal, hexagonal and orthorhombic systems with space groups R$\overline{3}$c~\cite{ramogayana2021density}, P$6_{3}$mc~\cite{Widmann_2024} and Pbnm~\cite{Hazen_1987}, respectively. Their experimental crystal parameters along with the values calculated by r$^2$SCAN and HSE06 are collected in Table~\ref{tab:lattice-parameters}. The calculated values are within 1\% of the reported experimental data for all three structures. In all materials, the \ce{Cr^3+} dopant substitutes an aluminium cation (\ce{Al^3+}), leading to the formation of six-coordinate dopant sites of various symmetries, as shown in Figure~\ref{fig:structures}. The geometries of the \ce{Cr^3+} dopant in each host, obtained with use of r$^2$SCAN and HSE06, are illustrated in Figure~\ref{fig:coordination} (all structures are given in Supplementary Information~\dag). The dopant site in \ce{AlB4O6N} is of high $O_{h}$ symmetry with minimal distortion from the perfect octahedral coordination characterized by bond length deviation not exceeding \SI{0.01}{\angstrom} and bond angles deviating from the perfect \SI{90}{\degree} by \SI{0.8}{\degree}. $\upalpha$-\ce{Al2O3} also has a single dopant site that exhibits $C_{3}$ symmetry and \ce{BeAl2O4} has two dopant sites characterized by $C_{s}$ (Wyckoff position 4\textit{c}) and $C_{i}$ (Wyckoff position 4\textit{a}) local symmetry point groups. In all cases, the geometries of the coordination spheres are close to that of a regular octahedron, with fairly small distortions resulting in symmetry lowering. In \ce{Cr^3+_{Al}}:$\upalpha$-\ce{Al2O3}, the distortion is characterized by a maximum deviation of \SI{0.05}{\angstrom} in \ce{Cr-O} bond lengths and a maximum deviation of \SI{10}{\degree} in bond angles, whereas in the case of \ce{Cr^3+_{Al}}:\ce{BeAl2O4} these values are \SI{0.09}{\angstrom} and \SI{10}{\degree}, respectively. Due to this, we use the notations of irreducible representations of the $O_{h}$ point group to label electronic states in all three systems and separately discuss splittings of the high-symmetry multiplets caused by the imperfections of dopant sites. As follows from the Tanabe-Sugano diagram of the $d^{3}$ configuration in an octahedral field, \ce{Cr^3+} cations must always have a high-spin quartet ($S=\nicefrac{3}{2}$) ground state ($^{4}\mathrm{A}_{2}$), and a set of low-spin quartet and doublet ($S=\nicefrac{1}{2}$) excited states, the order and energies of which depend on the interactions of the \ce{Cr^3+} dopant with its host. Transitions between these states correspond to the class of crystal field $d$-$d$ excitations and can be experimentally probed via ultraviolet--visible and photoluminescence spectroscopy. We further test the capabilities of DFT and pEHCF in reproducing the energies and spin-symmetries of excited $d$-multiplets of \ce{Cr^3+} dopants. \\

\begin{figure*}
\begin{centering}
\includegraphics[scale=0.5]{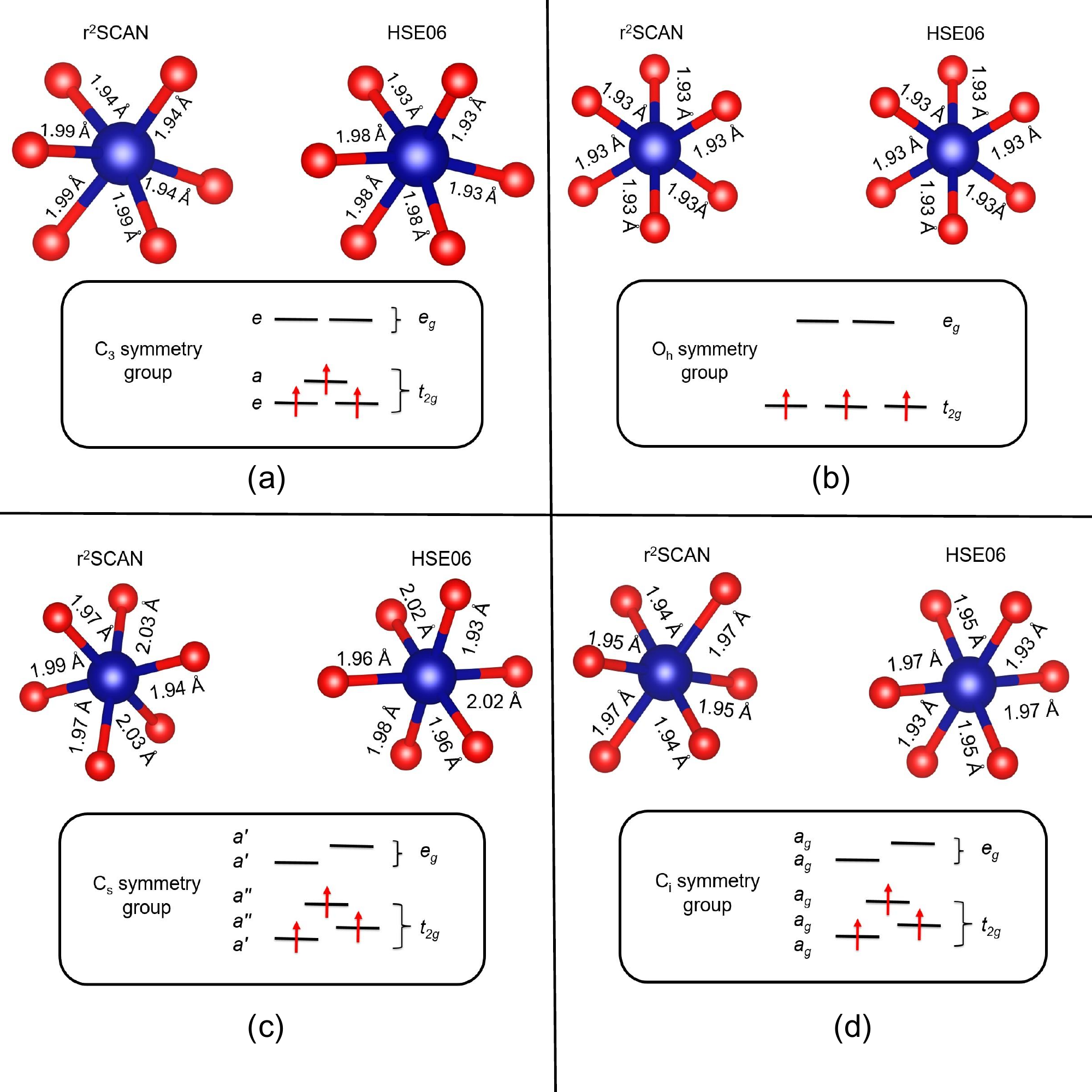}
\par\end{centering}
\caption{Ground-state geometries of the Cr coordination sphere and the splitting diagrams of the one-electron $d$-states for (a) $\upalpha$-Al$_2$O$_3$, (b) AlB$_4$O$_6$N, and BeAl$_2$O$_4$ at (c) $C_{s}$- and (d) $C_{i}$-sites. The corresponding splitting parameters of the one-electron $d$-states calculated with r$^{2}$SCAN, HSE06, and pEHCF are presented in Table \ref{tab:one-el}.}
\label{fig:coordination}
\end{figure*}

\begin{table*}
\caption{Experimental (Exp.) lattice parameters (in \si{\angstrom}) of $\upalpha$-\ce{Al2O3}, \ce{AlB4O6N}, and \ce{BeAl2O4} compared to the DFT calculations performed in this work.}
\centering{}
\renewcommand{\arraystretch}{1.5}
\begin{tabular}{cccccccccc}
\hline 
\multirow{2}{*}{} 
& \multicolumn{3}{c}{$\upalpha$-Al$_2$O$_3$} 
& \multicolumn{3}{c}{AlB$_4$O$_6$N} 
& \multicolumn{3}{c}{BeAl$_2$O$_4$} \tabularnewline
\cline{2-10}
& Exp.~\cite{Lucht_2003} & r$^2$SCAN & HSE06 
& Exp.~\cite{Widmann_2024} & r$^2$SCAN & HSE06 
& Exp.~\cite{Hazen_1987} & r$^2$SCAN & HSE06 \tabularnewline
$a$ & 4.76 & 4.76 & 4.75 & 5.03 & 5.03 & 5.02 & 9.42 & 9.40 & 9.39 \tabularnewline
$b$ & -- & -- & -- & -- & -- & -- & 5.48 & 5.47 & 5.47 \tabularnewline
$c$ & 12.98 & 12.98 & 12.96 & 8.23 & 8.23 & 8.22 & 4.43 & 4.42 & 4.41 \tabularnewline
\hline 
\end{tabular}
\label{tab:lattice-parameters}
\end{table*}

\begin{figure*}
\begin{centering}
\includegraphics[scale=0.55]{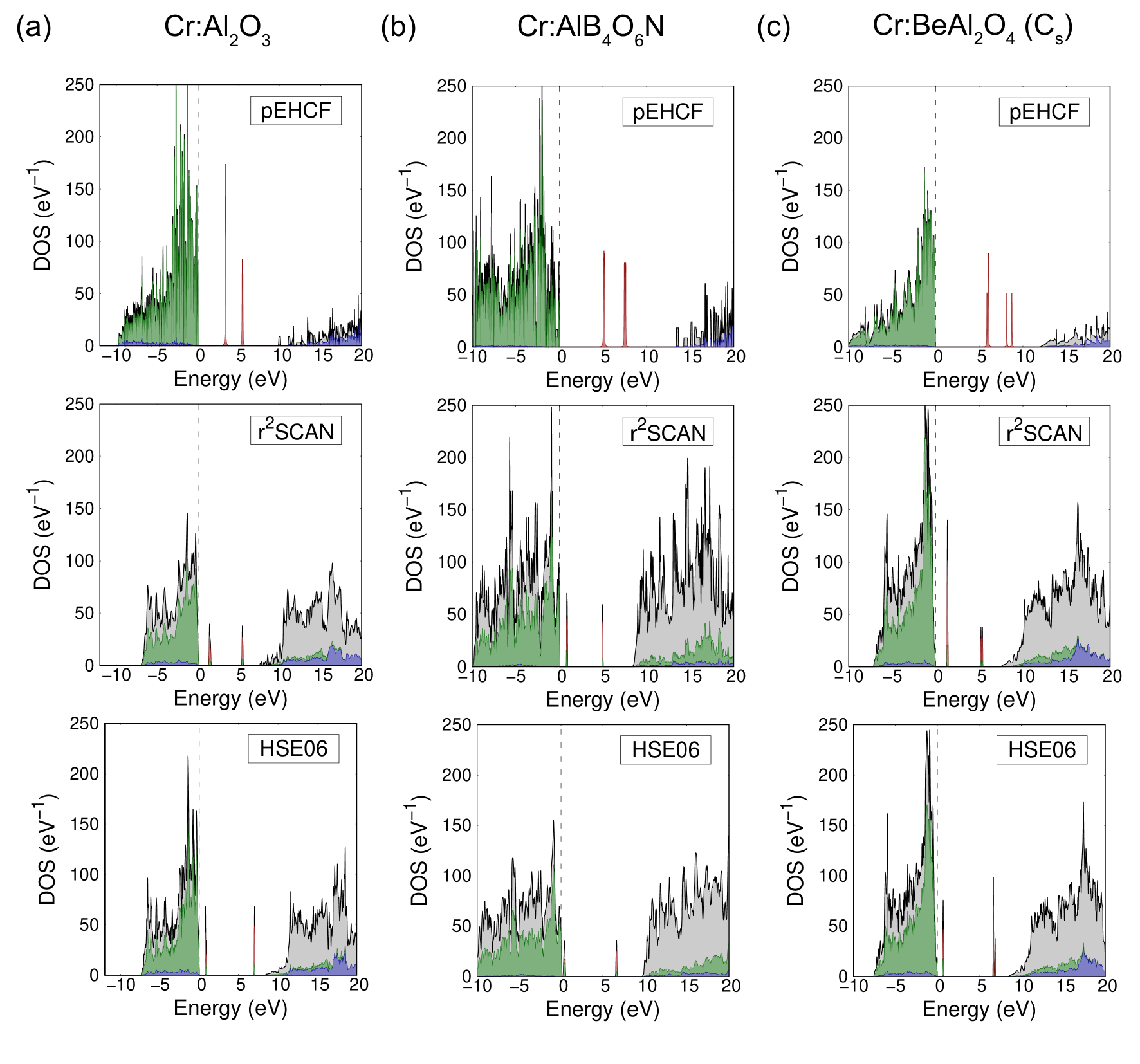}
\par\end{centering}
\caption{The atomic orbital-projected density of states (DOS) for \ce{Cr^3+}-doped (a) $\upalpha$-Al$_2$O$_3$, (b) \ce{AlB4O6N} and (c) \ce{BeAl2O4} ($C_{s}$), as calculated using the r$^2$SCAN, HSE06 and pEHCF. The total density of states, chromium 3$d$, oxygen 2$p$ and aluminium 3$p$ states are shown in gray, red, green and blue, respectively. Other atomic orbitals are not shown for clarity. For the purpose of comparison across all three methods, the position of the reference point on the energy axis is chosen such that the top of the $sp$-valence band corresponds to \SI{0}{\electronvolt}. In pEHCF, the peaks corresponding to the $d$-states indicate the position of the one-electron $d$-orbitals. This representation does not fully reflect the complexity of the electronic structure of the $d$-system containing multi-reference many-electron multiplets described by pEHCF.}
\label{fig:dos_cr_dop}
\end{figure*}

First, we analyse the one-electron states of the three systems, as calculated using DFT and pEHCF methods. The atomic orbital-projected density of states (DOS) for the ground states of \ce{Cr^3+}-doped $\upalpha$-\ce{Al2O3}, \ce{AlB4O6N}, and \ce{BeAl2O4} are shown in Figure~\ref{fig:dos_cr_dop}. For all systems, as can be seen from the r$^2$SCAN-calculated DOS, the narrow $d$-bands of \ce{Cr^3+} lie within the gap between the wide $sp$ valence and conduction bands of the host. The small width of the $d$-bands indicates a small degree of coupling between the $d$-shell and $sp$-states, supporting the assumption regarding the locality of $d$-shells that is employed within pEHCF. In the case of DFT,  there is a minor contribution from the oxygen 2$p$ states in the gap states reflecting a small degree of hybridization with the chromium 3$d$ states. The pEHCF-calculated DOS plots qualitatively agree with DFT regarding the position of the one-electron $d$-states and the structure of the frontier $sp$-bands. The valence band of the $sp$-subsystem mostly comprises oxygen $2p$ orbitals for each host material, whereas the conduction band has significant contributions from aluminium $3p$ orbitals. If the concentration of \ce{Cr^3+} dopants is sufficiently low, the gap between the valence and conduction $sp$-bands should be close to the band gap of the host material. 

Our calculations show that for the given unit cells, the $sp$-band gaps of the doped materials are smaller than the band gaps of the pure hosts by 0.1--\SI{0.2}{\electronvolt}, which can be compared against corresponding experimental values. Experimental values of the band gap of $\upalpha$-\ce{Al2O3} have been reported to range from 8.15--\SI{9.40}{\electronvolt} depending on conditions~\cite{french1990electronic}, while the reported experimental value for \ce{BeAl2O4} is \SI{9.00}{\electronvolt}~\cite{ivanov2005electronic}. The calculated values of the band gap are compared to the available experimental data in Table~\ref{tab:one-el}. It shows that pEHCF systematically overestimates the gap by 0.50--\SI{1.50}{\electronvolt} for \ce{Cr^3+_{Al}}:$\upalpha$-\ce{Al2O3} and by \SI{2.00}{\electronvolt} for \ce{Cr^3+_{Al}}:\ce{BeAl2O4} as the Hartree-Fock method is used to calculate the electronic structure of the $sp$-subsystem. In contrast, r$^2$SCAN underestimates the band gap by about \SI{1.5}{\electronvolt} for both materials whilst HSE06 gives the improved values of \SI{8.15}{\electronvolt} for $\upalpha$-\ce{Al2O3} and \SI{8.38}{\electronvolt} for \ce{BeAl2O4}. No experimental band gap value has been reported for AlB$_4$O$_6$N, but we calculate a band gap of \SI{9.78}{\electronvolt} using HSE06. Other reported theoretical band gap values range from 7.32--\SI{9.31}{\electronvolt}, depending on the computational method ~\cite{Widmann_2024}.

Splitting diagrams for the one-electron $d$-states and the ground state, as calculated with pEHCF, r$^2$SCAN and HSE06, are shown in Figure~\ref{fig:coordination}. All methods qualitatively follow the same symmetry considerations, exhibiting a significant $t_{2g}$–$e_g$ splitting characteristic of an octahedral crystal field. An additional minor splitting within the threefold degenerate $t_{2g}$ manifold is present in \ce{Cr^3+_{Al}}:$\upalpha$-Al$_2$O$_3$ ($C_3$ symmetry), while full degeneracy lifting occurs for \ce{Cr^3+_{Al}}:BeAl$_2$O$_4$, which possesses lower-symmetry ($C_s$ and $C_i$) dopant sites. Quantitatively, the absolute magnitudes of crystal-field splittings differ significantly between pEHCF and the DFT methods (Table~\ref{tab:2}). For instance, the r$^2$SCAN splitting of $t_{2g}$–$e_g$ has values of around 4.00--\SI{4.40}{\electronvolt}; HSE06 provides splittings of 6.10--\SI{6.60}{\electronvolt}, while pEHCF shows splittings of 2.08--\SI{3.47}{\electronvolt}. This disparity is to be expected due to the nature of the one-electron energies in both methods. pEHCF splitting diagrams correspond to the eigenvalues of the one-electron part of the effective Hamiltonian, whereas energy levels produced by DFT already include an effect of $d$-$d$ electron interactions. From this point of view, pEHCF produces splitting parameters that are usually discussed in the literature related to spectroscopy of transition metal ions, such as $10Dq$ in the $O_{h}$ crystal field. 

Many-electron multiplet energies calculated using r$^2$SCAN, HSE06, and pEHCF are detailed in Tables~\ref{tab:spectrum-Al2O3}, \ref{tab:spectrum-AlB4O6N}, and \ref{tab:spectrum-BeAl2O4} for the three materials, where they are compared against experimental values. Theoretically, the simplest transition is $^{4}\mathrm{A}_{2}\rightarrow{}^{4}\mathrm{T}_{2}$, which corresponds to the promotion of one electron from the $t_{2g}$-orbital to the $e_{g}$-orbital. Both multiplets can be accurately described using a single determinant wavefunction~\cite{Tanabe_1954}; the effect of static correlations should therefore be minor. r$^2$SCAN and HSE06 accurately reproduce the energy of this transition for all materials with absolute errors compared to experiment ranging between 0.10--\SI{0.15}{\electronvolt}, and pEHCF also provides the same level of accuracy. As shown in Table~\ref{tab:spectrum-Al2O3}, CASSCF calculations previously performed~\cite{Shang_2022} for a finite \ce{[CrO6]^9-} cluster cut out from \ce{Cr^3+}:\ce{Al2O3} give slightly larger errors for the $^{4}\mathrm{A}_{2}\rightarrow{}^{4}\mathrm{T}_{2}$ transition as compared to DFT and pEHCF. 

Other excited states, such as $^{2}\mathrm{E}$, have significant multi-reference character, and correct description of their electronic structure is therefore much more challenging for DFT. As has been previously shown with $\Delta$SCF, the low-spin excited states may not be achievable or could be significantly underestimated with respect to experimental values~\cite{thiering2017ab}. This is clearly illustrated by the results for the $^{2}\mathrm{E}$ multiplet that presents an interest for \ce{Cr^3+}-based spin qubits as it plays an important role in the inter-system crossing pathway~\cite{thiering2017ab}. Both r$^2$SCAN and HSE06 consistently underestimate the energy of the $^{2}\mathrm{E}$ state as compared to experiments, with absolute errors ranging between 0.50--\SI{0.60}{\electronvolt}. At the same time, pEHCF provides good accuracy for the $^{4}\mathrm{A}_2\rightarrow^{2}\mathrm{E}$ lines for all materials, with absolute errors ranging between 0.07--\SI{0.11}{\electronvolt}. The $^{2}\mathrm{T}_{1}$ and $^{2}\mathrm{T}_{2}$ states are even more complicated, as their energies cannot be calculated using $\Delta$SCF at all due to their multi-reference nature. This is because their many-determinant wavefunctions cannot be approximated by the one-electron population matrix that is used to set up a trial Kohn-Sham wavefunction in $\Delta$SCF. Full configuration interaction treatment of the $d$-shell, as implemented in pEHCF, permits the $^{2}\mathrm{T}_{1}$ and $^{2}\mathrm{T}_{2}$ wavefunctions to be determined and results in energies that are in good agreement with experimental data for all systems.

Regarding the second excited quartet state, $^{4}\mathrm{T}_{1}$, all methods agree well with experimental data. For \ce{Cr^3+_{Al}}:$\upalpha$-\ce{Al2O3}, pEHCF successfully captures the reported experimental values ranging from 3.01--\SI{3.12}{\electronvolt}~\cite{Cronemeyer_1966, Kusuma_2019, Song_2005}. r$^2$SCAN and HSE06 give excitation energies of \SI{3.31}{\electronvolt} and \SI{3.05}{\electronvolt}, respectively. For \ce{Cr^3+_{Al}}:\ce{AlB4O6N}, the second excited quartet is calculated to be \SI{3.30}{\electronvolt} in pEHCF and 3.46--\SI{3.49}{\electronvolt} by DFT methods, which can be compared to the experimental value of \SI{3.26}{\electronvolt}~\cite{Widmann_2024}. Finally, for \ce{Cr^3+_{Al}}:\ce{BeAl2O4}, the experimentally reported value of \SI{3.02}{\electronvolt}~\cite{Demirbas_2019} agrees well with our r$^2$SCAN-calculated energy of \SI{2.98}{\electronvolt} for the $C_{s}$-symmetrized site and our HSE06-calculated energy of \SI{3.02}{\electronvolt} for the $C_{i}$-symmetrized site. In contrast, pEHCF overestimates the energy of $^{4}\mathrm{A}_{2} \rightarrow{}^{4}\mathrm{T}_{1}$ transition by \SI{0.20}{\electronvolt} and \SI{0.10}{\electronvolt} for the $C_{s}$- and $C_{i}$-symmetrized sites, respectively.

We also note an interesting discrepancy with experimental data for \ce{Cr^3+_{Al}}:\ce{BeAl2O4} at the $C_{i}$ dopant site. As shown in Table~\ref{tab:spectrum-BeAl2O4}, the pEHCF-, r$^2$SCAN-, and HSE06-calculated energies for the $^{4}\mathrm{A}_{2}\rightarrow{}^{4}\mathrm{T}_{2}$ transition are in good agreement with the experimentally observed line for the $C_{s}$ dopant site. However, for the $C_{i}$ dopant site, the $^{4}\mathrm{A}_{2}\rightarrow{}^{4}\mathrm{T}_{2}$ transition is significantly overestimated by all computational methods: by around 0.6--\SI{0.9}{\electronvolt} with pEHCF and \SI{0.4}{\electronvolt} with DFT. This indicates that the $C_{s}$ dopant site might be largely responsible for the experimental emission, whereas the $C_{i}$ dopant site plays a minor role. This conclusion is further supported by our DFT calculations showing that \ce{Cr^3+} in the $C_s$-symmetrized site lies lower in energy than the $C_{i}$-symmetrized site by \SI{0.17}{\electronvolt} and \SI{0.19}{\electronvolt} for r$^2$SCAN and HSE06, respectively. A wide range of experimental studies confirm the preference of the $C_{s}$ site. For instance, electron paramagnetic resonance and optical absorption spectroscopic analyses of \ce{BeAl2O4} indicate that 75\% of \ce{Cr^3+} ions substitute the $C_{s}$-symmetrized site~\cite{trindade2011temperature}. Using X-ray absorption spectroscopy, ~\citet{bordage2012site} also confirmed \ce{Cr^3+} substitution at the $C_{s}$-symmetrized site to be 70\%. These experimental findings are consistent with our computational results suggesting that the optical features observed experimentally predominantly correspond to Cr$^{3+}$ substitution at the energetically and structurally favoured $C_{s}$-symmetrized site.

Our results show that, for the excited multiplets of \ce{Cr^3+},  r$^2$SCAN and HSE06 functionals consistently give close energy values and 
perform particularly well for the high-spin states. Previous work on NV-like defects has shown that both hybrid density-functional approximation and meta-GGA can yield reliable predictions for the formation energy and charge transition levels~\cite{10.1063/5.0252129}.  However, as shown in Ref.~\cite{PhysRevB.87.205201}, HSE06 may fail for transition metal containing systems due to the Coulombic interactions between localised $d$-electrons not being sufficiently screened. This results in an incomplete removal of self-interaction error and violation of the generalised Koopmans' condition. This makes r$^2$SCAN a competitive, lower-cost alternative to hybrid density-functional approximations for systems containing transition metal atoms, particularly when combined with the pEHCF multi-reference treatment of the $d$-shell.

%
%
%

\begin{table*}
\small
\caption{\label{tab:one-el}Energies of the one-electron $d$-states of \ce{Cr^3+} dopants and $sp$-band gaps calculated using different quantum mechanical methods. The calculated values of the band gap are compared to available experimental data for $\upalpha$-Al\textsubscript{2}O\textsubscript{3} and BeAl\textsubscript{2}O\textsubscript{4}.}

\begin{centering}
\resizebox{\textwidth}{!}{%
\renewcommand{\arraystretch}{1.5}
\begin{tabular}{ccccccccccccc}
\hline 
host $\rightarrow$ & \multicolumn{3}{c}{$\upalpha$-Al\textsubscript{2}O\textsubscript{3}} & \multicolumn{3}{c}{AlB\textsubscript{4}O\textsubscript{6}N} & \multicolumn{3}{c}{BeAl\textsubscript{2}O\textsubscript{4} ($C_{s}$)} & \multicolumn{3}{c}{BeAl\textsubscript{2}O\textsubscript{4} ($C_{i}$)}\\
\cline{2-13}
 & pEHCF & r$^2$SCAN & HSE06 & pEHCF & r$^2$SCAN & HSE06 & pEHCF & r$^2$SCAN & HSE06 & pEHCF & r$^2$SCAN & HSE06 \\
\hline 
\multirow{3}{*}{$t_{2g}$ [\si{\electronvolt}]} & 0.00 & 0.00 & 0.00 & 0.00 & 0.00 & 0.00 & 0.00 & 0.00 & 0.00 & 0.00 & 0.00 & 0.00 \\
 & 0.00 & 0.00 & 0.00 & 0.06 & 0.05 & 0.11 & 0.19 & 0.05 & 0.05 & 0.14 & 0.09 & 0.08 \\
 & 0.03 & 0.10 & 0.10 & 0.10 & 0.05 & 0.11 & 0.30 & 0.07 & 0.09 & 0.37 & 0.19 & 0.15 \\
\hline 
\multirow{2}{*}{$e_{g}$ [\si{\electronvolt}]} & 2.08 & 4.04 & 6.21 & 2.40 & 4.15 & 6.32 & 2.50 & 3.92 & 6.11 & 2.95 & 4.29 & 6.47 \\
 & 2.15 & 4.04 & 6.21 & 2.55 & 4.15 & 6.32 & 2.69 & 4.06 & 6.26 & 3.17 & 4.40 & 6.59 \\
\hline 
$E_\mathrm{gap}^{sp}$ [\si{\electronvolt}] & 9.86 & 7.22 & 8.21 & 13.45 & 7.99 & 9.72 & 11.71 & 7.45 & 8.45 & 11.77 & 7.45 & 8.42 \\
Exp. $E_\mathrm{gap}^{sp}$ [\si{\electronvolt}] &  & 8.15--9.40~\cite{french1990electronic} &  &  & - &  &  & 9.00~\cite{ivanov2005electronic} &  &  & 9.00~\cite{ivanov2005electronic} &  \\
\hline 
\end{tabular}}
\par\end{centering}
\label{tab:2}
\end{table*}

\begin{table*}
\begin{centering}
\caption{\label{tab:spectrum-Al2O3}Calculated energies (in \si{\electronvolt}) of excited $d$-multiplets of \ce{Cr^3+_{Al}}:$\upalpha$-\ce{Al2O3} using different quantum mechanical methods compared to optical absorption spectroscopy data and CASSCF calculations for a \ce{[CrO6]^9-} cluster (cut from \ce{Cr^3+_{Al}}:$\upalpha$-\ce{Al2O3}) from ~\citet{Shang_2022}. The $^{4}\mathrm{A}_{2}\rightarrow{}^{2}\mathrm{E}$ transition is extracted from the \textit{R}-line in the experimental references.}
\par\end{centering}
\centering{}
\renewcommand{\arraystretch}{1.5}
\begin{tabular}{cccccc}
\hline 
Transition & Experiment & pEHCF & r$^2$SCAN & HSE06 & CASSCF\tabularnewline
\hline 
$^{4}\mathrm{A}_{2}\rightarrow{}^{2}\mathrm{E}$ & 1.78~\cite{Cronemeyer_1966}; 1.80~\cite{Kusuma_2019} & 1.90 & 1.21 & 1.14 & 1.78~\cite{Shang_2022}\tabularnewline
$^{4}\mathrm{A}_{2}\rightarrow{}^{2}\mathrm{T}_{1}$ & not resolved & 1.97; 2.00 & - & - & -\tabularnewline
$^{4}\mathrm{A}_{2}\rightarrow{}^{4}\mathrm{T}_{2}$ & 2.22~\cite{Cronemeyer_1966,Kusuma_2019}; 2.25~\cite{Song_2005}; 2.28~\cite{Cronemeyer_1966} & 2.08; 2.14 & 2.37 & 2.36 & 2.49~\cite{Shang_2022}\tabularnewline
$^{4}\mathrm{A}_{2}\rightarrow{}^{2}\mathrm{T}_{2}$ & not resolved & 2.82; 2.83; 2.90 & - & - & -\tabularnewline
$^{4}\mathrm{A}_{2}\rightarrow{}^{4}\mathrm{T}_{1}$ & 3.01-3.03~\cite{Cronemeyer_1966,Kusuma_2019,Song_2005}; 3.12~\cite{Cronemeyer_1966} & 2.94; 2.97; 3.11 & 3.31 & 3.05 & -\tabularnewline
\hline 
\end{tabular}
\end{table*}

\begin{table*}
\begin{centering}
\caption{\label{tab:spectrum-AlB4O6N}Calculated energies (in \si{\electronvolt}) of excited $d$-multiplets of \ce{Cr^3+_{Al}}:\ce{AlB4O6N} using different quantum mechanical methods compared to photoluminescence excitation and emission spectroscopy data for the $^{4}\mathrm{A}_{2}\rightarrow{}^{2}\mathrm{E}$ transition from ~\citet{Widmann_2024}.}
\par\end{centering}
\centering{}
\renewcommand{\arraystretch}{1.5}
\begin{tabular}{ccccc}
\hline 
Transition & Experiment & pEHCF & r$^2$SCAN & HSE06\tabularnewline
\hline 
$^{4}\mathrm{A}_{2}\rightarrow{}^{2}\mathrm{E}$ & 1.81 & 1.91 & 1.25 & 1.12\tabularnewline
$^{4}\mathrm{A}_{2}\rightarrow{}^{2}\mathrm{T}_{1}$ & 1.89 & 1.99; 2.01 & - & -\tabularnewline
$^{4}\mathrm{A}_{2}\rightarrow{}^{4}\mathrm{T}_{2}$ & 2.43 & 2.40; 2.42; 2.44 & 2.46 & 2.49\tabularnewline
$^{4}\mathrm{A}_{2}\rightarrow{}^{2}\mathrm{T}_{2}$ & 2.70 & 2.89; 2.92 & - & -\tabularnewline
$^{4}\mathrm{A}_{2}\rightarrow{}^{4}\mathrm{T}_{1}$ & 3.26 & 3.30; 3.32 & 3.49 & 3.46\tabularnewline
\hline 
\end{tabular}
\end{table*}

\begin{table*}
\begin{centering}
\caption{\label{tab:spectrum-BeAl2O4}Calculated energies (in \si{\electronvolt}) of excited $d$-multiplets of \ce{Cr^3+_{Al}}:\ce{BeAl2O4} using different quantum mechanical methods compared to ultraviolet–visible spectroscopy data from ~\citet{Demirbas_2019}.}
\par\end{centering}
\centering{}
\renewcommand{\arraystretch}{1.5}
\begin{tabular}{cccccccc}
\hline 
\multirow{2}{*}{Transition} & \multirow{2}{*}{Experiment} & \multicolumn{3}{c}{$C_{s}$ site} & \multicolumn{3}{c}{$C_{i}$ site}\tabularnewline
\cline{3-8} \cline{4-8} \cline{5-8} \cline{6-8} \cline{7-8} \cline{8-8} 
 & & pEHCF & r$^2$SCAN & HSE06 & pEHCF & r$^2$SCAN & HSE06\tabularnewline
\hline 
$^{4}\mathrm{A}_{2}\rightarrow{}^{2}\mathrm{E}$ & 1.82 & 1.89; 1.90 & 1.27 & 1.17 & 1.80; 1.89 & 1.22 & 1.15\tabularnewline
$^{4}\mathrm{A}_{2}\rightarrow{}^{2}\mathrm{T}_{1}$ & 1.91 & 1.95; 1.99; 2.00 & - & - & 1.91; 2.01; 2.04 & - & -\tabularnewline
$^{4}\mathrm{A}_{2}\rightarrow{}^{4}\mathrm{T}_{2}$ & 2.10 & 2.34; 2.41; 2.52 & 2.18 & 2.16 & 2.72; 2.91; 2.95 & 2.47 & 2.49\tabularnewline
$^{4}\mathrm{A}_{2}\rightarrow{}^{2}\mathrm{T}_{2}$ & - & 2.91; 2.95; 2.96 & - & - & 3.00; 3.04; 3.09 & - & -\tabularnewline
$^{4}\mathrm{A}_{2}\rightarrow{}^{4}\mathrm{T}_{1}$ & 3.02 & 3.23; 3.37 & 2.98 & 2.72 & 3.64 & 3.16 & 3.02\tabularnewline
\hline 
\end{tabular}
\end{table*}

\section*{Conclusions}

In this work, we have used the r$^2$SCAN meta-generalized gradient approximation and the range-separated HSE06 hybrid density-functional approximation, alongside the wavefunction-based pEHCF method, to investigate the electronic structure of the low-lying $d$-$d$ excited states of chromium(III) dopants in three wide band gap materials. Our results demonstrate that the energy of the $^{4}\mathrm{A}_{2}\rightarrow{}^{4}\mathrm{T}_{2}$ and $^{4}\mathrm{A}_{2}\rightarrow{}^{4}\mathrm{T}_{1}$ transitions between single-reference quartet states can be accurately described by all three methods with a similar level of accuracy. r$^2$SCAN and HSE06 yield similar results for the high-spin excited states, making r$^2$SCAN an accurate and relatively low-cost density-functional approximation compared to HSE06. At the same time, high-spin to low-spin transitions are much more difficult to capture with DFT methods. However, as the pEHCF method treats open $d$-shells at the configuration interaction level, the energies of all the transitions can be calculated at a high level of accuracy. A combination of DFT and pEHCF, therefore, provides a reliable tool for the quantitative study of transition metal-doped materials. While DFT is effective in predicting ground-state geometry and properties and certain high-spin excited states, its underestimation of the energies of low-spin configurations due to their multi-reference nature is a considerable limitation. In contrast, pEHCF, with its full configuration interaction treatment of the $d$-shell, excels at capturing these multiplets, as well as the energies of spin-forbidden transitions. Our proposed computational framework can guide future materials design for optical, magnetic and quantum technologies.

\section*{Author contributions}

\noindent I.P. -- conceptualization, investigation (pEHCF calculations), data curation, visualization, writing - original draft \& editing. P.-P.F. -- investigation (DFT calculations), data curation, visualization, writing - original draft \& editing. S.C. -- investigation (DFT calculations), writing - review \& editing. A.L.T. -- conceptualization, writing - review \& editing. K.I. -- conceptualization, supervision, writing - review \& editing. E.B. -- conceptualization, supervision, writing - review \& editing, project administration.

\section*{Conflicts of interest}

There are no conflicts to declare.

\section*{Data availability}

The structural data supporting this article have been included as part of the Supplementary Information.\dag

\section*{Acknowledgements}

E.B. acknowledges a Royal Society Wolfson Fellowship and the EPSRC Program Grant "Enabling Net Zero and the AI Revolution with ultra-low energy 2D Materials and Devices (NEED2D)" [UKRI-1249] for funding. E.B. and I.P. are funded by the EPSRC Program Grant “Metal Atoms on Surfaces and Interfaces (MASI) for Sustainable Future” [EP/V000055/1]. K.I. and P.-P.F. are funded by the EPSRC Fellowship Programme [EP/W028131/1] and S.C. is funded by Wellcome Leap as part of the Quantum for Bio Program. The work of A.L.T. is supported by the Ministry of Science and Higher Education of the Russian Federation. Computing resources were provided by the University of Nottingham, the EPSRC-funded HPC Midlands+ consortium [EP/T022108/1] for access to Sulis, and the EPSRC-funded High-End Computing Materials Chemistry Consortium [EP/X035859/1] for access to the ARCHER2 UK National Supercomputing Service (\url{https://www.archer2.ac.uk}). P.-P.F. acknowledges Prof. {\'A}d{\'a}m Gali and Dr. Gerg{\H o} Thiering (Wigner Research Centre for Physics, Hungarian Academy of Sciences) for useful discussions. The authors also thank the reviewers for their valuable assessment of this work.




\renewcommand\refname{References}

\bibliography{rsc} 
\bibliographystyle{rsc} 
\end{document}


\onehalfspacing
\date{}
\maketitle

\section*{Structures of all systems considered in the manuscript optimised with r$^{2}$SCAN and HSE06 functionals}

Here, we provide crystal structures of four system considered in this work: Cr$^{3+}$:Al$_{2}$O$_{3}$, Cr$^{3+}$:AlB$_{4}$O$_{6}$N, Cr$^{3+}$:BeAl$_{2}$O$_{4}$ ($C_{s}$) and Cr$^{3+}$:BeAl$_{2}$O$_{4}$ ($C_{i}$) optimised with two DFT functionals: r$^{2}$SCAN and HSE06. First line of every structure specifies the type of the system and the functional that was used for geometry optimisation.

\newpage
\pagenumbering{gobble}

\lstset{basicstyle=\footnotesize}

\begin{lstlisting}
Cr:AL2O3 optimised with r2SCAN                                
1.0000000000000000     
 9.5099287033000000    0.0000000000000000    0.0000000000000000
-4.7549643517000000    8.2358398451999992    0.0000000000000000
 0.0000000000000000    0.0000000000000000   12.9827966690000007
Cr   Al    O             
1    47    72
Direct
     0.5000001334799029    0.5000002060993880    0.3491658111768497
     0.1666382845128986    0.3332791511453660    0.8147808971962771
     0.1666668259185974    0.8333335626504534    0.8149634203037339
     0.6667210394320682    0.3333592532280202    0.8147809302800303
     0.6666413280818873    0.8333620856448546    0.8147807747231822
     0.3329294484196804    0.1666458289479147    0.6854671338135891
     0.3337164561589516    0.6670707083357994    0.6854670880491338
     0.8333335351957281    0.1666667845430753    0.6854924494926474
     0.8333545001093516    0.6662839467444658    0.6854670172558592
     0.9996405869598365    0.9996569395915997    0.6485983484311363
     0.0003432660246438    0.4999836945551231    0.6485983888365455
     0.5000163425890332    0.0003596498483208    0.6485983805369342
     0.5000000129094394    0.5000000214827923    0.6478005863559601
     0.1660883533869021    0.3324347007533339    0.5190661771441131
     0.1666665691865540    0.8333332848868704    0.5186680022057629
     0.6675652267393376    0.3336536133276167    0.5190661261228267
     0.6663463809200503    0.8339114556130499    0.5190661601145276
     0.9997562330526465    0.0000578150660551    0.1481765962682805
     0.9999419582231656    0.4996982089536166    0.1481765407958149
     0.5003015404745955    0.0002437271152725    0.1481766669450938
     0.5000000759138646    0.5000000958051576    0.1461986174342064
     0.1664454853980343    0.3333437559043645    0.0182759689228555
     0.1666667475895917    0.8333333827706166    0.0183848670668458
     0.6666562229191554    0.3331016203382222    0.0182759207444133
     0.6668983279878491    0.8335545844538782    0.0182760107146987
     0.3333525016589157    0.1666813346556285    0.9811234626705692
     0.3333288181534100    0.6666475102301546    0.9811234578429276
     0.8333333225567719    0.1666666875187539    0.9814677574049109
     0.8333187492345986    0.6666711799863339    0.9811234418975622
     0.9999892174709238    0.9999638994701923    0.8519649412382527
     0.0000364370545936    0.5000253149951490    0.8519650264026202
     0.4999747283867703    0.0000111384856704    0.8519650331768451
     0.5000000191853406    0.5000000054100775    0.8515493462378285
     0.3330504252102049    0.1653662884487862    0.4821704395836514
     0.3323158600277389    0.6669493577526494    0.4821703680452459
     0.8333335497109948    0.1666668832405612    0.4818124941316527
     0.8346335616108369    0.6676840895507790    0.4821703045797556
     0.9999022035655817    0.9997088128188333    0.3520084614233296
     0.0002908903856060    0.5001930275554183    0.3520083253368256
     0.4998065065431904    0.0000976825280141    0.3520085899634681
     0.1648732254557401    0.3319500829746334    0.3153243663809686
     0.1666664455906660    0.8333331575238353    0.3151282879814694
     0.6680496355768925    0.3329229329356003    0.3153243260477842
     0.6670768357663719    0.8351268098354712    0.3153243533231157
     0.3322376239259620    0.1651107164682494    0.1844495202033644
     0.3328728938266816    0.6677619836160360    0.1844495517523424
     0.8333332874325002    0.1666666494628155    0.1850746221682367
     0.8348889327858265    0.6671267289281794    0.1844495235244117
     0.1795389848580988    0.1658957573845661    0.5843930806519343
     0.1801809375245629    0.6667860978532810    0.5836333991688045
     0.6802952238284964    0.1668685960116154    0.5837715415568578
     0.6802942978796944    0.6667015857329035    0.5832871947766811
     0.3468505040738089    0.3469879201037931    0.7500117260470400
     0.3468067305470054    0.8469675118754120    0.7503927757645518
     0.8470322333682889    0.3468057929104083    0.7503468639189887
     0.8468005830462471    0.8466687240149189    0.7501287335407538
     0.4864073555551370    0.3197057734713741    0.5832872026968113
     0.4863568601244934    0.8204611331777950    0.5843930644942082
     0.9865734494048685    0.3197048837415338    0.5837715648554014
     0.9866052791233620    0.8198190267154284    0.5836332651747784
     0.3332139093657487    0.0133947862808930    0.5836334194696871
     0.3332984335655920    0.5135927156180251    0.5832872192071077
     0.8331312847269045    0.0134266530555716    0.5837714830853352
     0.8341042875796342    0.5136432437394285    0.5843931076561536
     0.9997737572056861    0.1529682185979427    0.7503468918886771
     0.0001608603595409    0.6531934367637571    0.7503927669540493
     0.4998679100200029    0.1531996185539712    0.7501289041904213
     0.5001376799302467    0.6531498461586390    0.7500117167551019
     0.1530329728299480    0.9998394603176911    0.7503928086948556
     0.1533318666782981    0.5001324069838314    0.7501289177202769
     0.6531943390644742    0.0002265546210688    0.7503468836729853
     0.6530124273306238    0.4998628949266412    0.7500117484151306
     0.0135002687526071    0.3333144799351038    0.9166740368581574
     0.0135405240066185    0.8332958878523534    0.9166977113869987
     0.5135373561551724    0.3333524635797983    0.9164842981178044
     0.5135282726082935    0.8333854995412229    0.9165864297601453
     0.1799307538312920    0.0133312768462163    0.0832402350053970
     0.1798844569331328    0.5137179508133337    0.0828676303684909
     0.6802535181481940    0.0134627418594747    0.0832628663272320
     0.6805278406889921    0.5137211502073172    0.0827200034293599
     0.3198151960852212    0.4864627804041033    0.9164843330344263
     0.3197553898795034    0.9864594833073622    0.9166977739531617
     0.8198573302163523    0.4864718417736602    0.9165864399617474
     0.8198142421808616    0.9864997108176951    0.9166740060994234
     0.1666145854423497    0.1801428233012338    0.9165864812567334
     0.1667041781160453    0.6802447418141151    0.9166977721447603
     0.6666855476899159    0.1801857920536733    0.9166740382430001
     0.6666477447056308    0.6801850979214978    0.9164841740404402
     0.3331932925280289    0.3194720684519216    0.0827201427539379
     0.3334005306487036    0.8200691411762813    0.0832402783091473
     0.8332092140069900    0.3197462637831652    0.0832627593845793
     0.8338335297735211    0.8201155143962015    0.0828675487259865
     0.4862820066486139    0.1661664707379102    0.0828676823986056
     0.4862787447801130    0.6668066790724224    0.0827201095787460
     0.9865371107959405    0.1667907005839287    0.0832628580293894
     0.9866686391985484    0.6665994318042686    0.0832401795060909
     0.3466834304050895    0.9997286716282281    0.2499396184651214
     0.3427787207761722    0.4981675097619344    0.2473072495507793
     0.8471281551410558    0.0003737759520964    0.2498854294637044
     0.8470840161097074    0.5001561931608406    0.2498431179930455
     0.0117986955100188    0.1790092941225106    0.4167529825225185
     0.0141851446040843    0.6803417983163504    0.4169491437482212
     0.5135204423226185    0.1803864818448527    0.4165953726632552
     0.5141659209157974    0.6873937106493841    0.4182531990621215
     0.1530721779126259    0.1529155959511615    0.2498433187474905
     0.1530452724324695    0.6533164275694562    0.2499395294120761
     0.6532453827811622    0.1528716271036629    0.2498855656404398
     0.6553887724461428    0.6572215260785815    0.2473071665531540
     0.9996262855680285    0.3467543163067291    0.2498855088426060
     0.0002710584231126    0.8469545562385146    0.2499395629197347
     0.5018322867390501    0.3446110841339401    0.2473072741012542
     0.4998432221648723    0.8469276800727160    0.2498433695018192
     0.1668661839167551    0.4864793862747826    0.4165953144399678
     0.1661565822803845    0.9858145999390835    0.4169491934536415
     0.6732274251694378    0.4858337203645858    0.4182531695382872
     0.6672106701460973    0.9882013091439086    0.4167529645153218
     0.3126062973663047    0.3267722499592098    0.4182532014216085
     0.3196579844809218    0.8338432518595694    0.4169492396254167
     0.8209905085242308    0.3327892355953427    0.4167529729015952
     0.8196132333526203    0.8331338791307510    0.4165953316379603
\end{lstlisting}
\newpage

\begin{lstlisting}
Cr:Al2O3 optimised with HSE06                
1.0000000000000000     
 9.5053443908999995    0.0000000000000000    0.0000000000000000
-4.7526721953999997    8.2318697142000001    0.0000000000000000
 0.0000000000000000    0.0000000000000000   12.9672012329000008
Cr   Al    O             
1    47    72
Direct
  0.1666622046383495  0.3333284387352009  0.8176103682872125
  0.1668722448700848  0.8335955124827379  0.8145406851295007
  0.6667222174266314  0.3331265305104196  0.8145412611542326
  0.6664048187248071  0.8332780423828368  0.8145395099749848
  0.3337366364499132  0.1658012879885860  0.6846500332428533
  0.3341986201913656  0.6679368525175775  0.6846495583952361
  0.8320670054937551  0.1662619672971317  0.6846495552459118
  0.8333311578573870  0.6666646499254441  0.6850270680557671
  0.9992141543314048  0.9995337638445960  0.6474956085583727
  0.0003217412106764  0.5007868373496649  0.6474930806568011
  0.4999978332309070  0.9999984615026563  0.6479077524126211
  0.5004652095081212  0.4996805833692122  0.6474927255454830
  0.1666652765114733  0.3333310909675404  0.5188374019364375
  0.1666566559229921  0.8335702057574395  0.5181827236193044
  0.6669104446332312  0.3333422741477037  0.5181802048985347
  0.6664291656622652  0.8330846969162238  0.5181821804479512
  0.0000010986270595  0.9997504607034671  0.1482543012674213
  0.9997514298865156  0.4999997059408017  0.1482530987447106
  0.5000004838544925  0.9999984842357037  0.1481738159256452
  0.5002508038341347  0.5002510190490170  0.1482536451798566
  0.1666657591378282  0.3333324042413253  0.0207915918274324
  0.1664213618561661  0.8332587828720861  0.0186181478686009
  0.6668378203270109  0.3335792685535779  0.0186175549739289
  0.6667413499262480  0.8331609276929512  0.0186173242612924
  0.3342566218054159  0.1661776327708111  0.9820455588786601
  0.3338238396938991  0.6680815482001421  0.9820456320026949
  0.8319195505837840  0.1657407084848330  0.9820456796200290
  0.8333327557256283  0.6666666133382506  0.9815687387831034
  0.9985599375039484  0.9980252868878381  0.8514004215614932
  0.9994634304470651  0.5014392205546017  0.8513997111040368
  0.4999988177014103  0.9999994698568742  0.8517093125697457
  0.5019748407501581  0.5005361150613226  0.8513997613013942
  0.3336430430576982  0.1669443526752872  0.4811270746414422
  0.3330562096655427  0.6666973911055223  0.4811267968109263
  0.8333048592650627  0.1663581518782706  0.4811259534704959
  0.8333336234082793  0.6666665804764733  0.4811649208059237
  0.9999505026267457  0.9999727526319120  0.3520061813717774
  0.0000224549966674  0.5000510656441861  0.3520064093074140
  0.5000011020246760  0.0000020879340070  0.3518755751738638
  0.5000337240876789  0.4999814983195066  0.3520069112300064
  0.1666663840445892  0.3333328595264362  0.3149309373842897
  0.1666751813808602  0.8333353056707367  0.3145799734459942
  0.6666582222436972  0.3333242104009386  0.3145793041076743
  0.6666651572231075  0.8333409200535087  0.3145784032282606
  0.3333153186981548  0.1665974843554210  0.1856878731412124
  0.3334017691369198  0.6667191839830622  0.1856863371952073
  0.8332837104794919  0.1666831696478752  0.1856869457258838
  0.8333345532188190  0.6666696733885260  0.1853935319112594
  0.1803281975044939  0.1666847690337576  0.5833375583248710
  0.1802976017362568  0.6673386081700059  0.5823425706341681
  0.6801007069164555  0.1665821558824803  0.5830872132708933
  0.6801057121786016  0.6664992944638044  0.5829657634976115
  0.3536592203101208  0.3475987607212545  0.7484185377924746
  0.3470145649152414  0.8475238172849089  0.7497251259684887
  0.8470485883651548  0.3468657537971396  0.7500506860223624
  0.8458173436722873  0.8452217263460255  0.7499382495470357
  0.4870409303402354  0.3197027644317458  0.5823430812466128
  0.4864823717533895  0.8199003330404935  0.5830872561940907
  0.9863569332770865  0.3196727287048446  0.5833375877844631
  0.9863937708158730  0.8198916825319316  0.5829656941776804
  0.3334184787795778  0.0135183928559570  0.5830875749315396
  0.3333142775648099  0.5136426639443101  0.5833371218608079
  0.8326616390683199  0.0129586332972309  0.5823426779110932
  0.8334990790373098  0.5136048510433397  0.5829662573000221
  0.9939340820628360  0.1463373182254273  0.7484194526619348
  0.9994047220209623  0.6541801354105701  0.7499383350031152
  0.5005092693560371  0.1529849954710727  0.7497251714752053
  0.4998159456976197  0.6529508483793265  0.7500505871817467
  0.1531347489598431  0.0001823435080937  0.7500514353375820
  0.1524021778118509  0.5060652144742974  0.7484182450395593
  0.6524750186859762  0.9994923501041910  0.7497237470991820
  0.6547734916791867  0.5005919106085699  0.7499382005964463
  0.0115418312676638  0.3350795913109863  0.9190328359930149
  0.0134544549276399  0.8329296896277754  0.9167730988990641
  0.5135226023292958  0.3331933763488450  0.9168008583460718
  0.5136276330164762  0.8335829634482153  0.9167420652921976
  0.1804064638800753  0.0132850986775281  0.0837841450483197
  0.1804193155605347  0.5138365890575045  0.0838975432309255
  0.6800928015258378  0.0133709048404924  0.0834692909885035
  0.6801553700649734  0.5135807878501168  0.0834505723081378
  0.3235358855120793  0.4884583695327080  0.9190327882427241
  0.3199549356069582  0.9863722706828781  0.9167432753375095
  0.8194755873199995  0.4865448842090885  0.9167723909095784
  0.8196692511707724  0.9864758415370289  0.9168007325777907
  0.1649212014822297  0.1764641350710434  0.9190333451087653
  0.1668060649617900  0.6803293929091936  0.9168005242799921
  0.6664166453609894  0.1800446807034461  0.9167430141345605
  0.6670685284921163  0.6805231174155324  0.9167721766546819
  0.3334172516482852  0.3195809758472450  0.0838978223365459
  0.3332785714952422  0.8199080768295062  0.0834689257814176
  0.8328796138277710  0.3195955507217789  0.0837847623141883
  0.8334285501645340  0.8198461138526838  0.0834493473494788
  0.4866292695611847  0.1667217068299252  0.0834697743595640
  0.4867139988612550  0.6671183183606075  0.0837845777727608
  0.9861639651189265  0.1665835969638110  0.0838976840614407
  0.9864186288499042  0.6665719168025390  0.0834495549112262
  0.3469519729314499  0.0000062219330843  0.2499694148765883
  0.3468890677921266  0.4999601476976920  0.2501133951002785
  0.8468841823753124  0.9999663766522104  0.2500531411129288
  0.8469303898708276  0.5000063992130848  0.2500027900593977
  0.0136248590637962  0.1802274862225488  0.4165989401433663
  0.0134650999756616  0.6803367009653201  0.4163539820421605
  0.5135968420252155  0.1801453364008481  0.4163153263271724
  0.5133922455067648  0.6801710795467955  0.4165295174600345
  0.1530712384193436  0.1531146989064140  0.2501140544685825
  0.1530799522365740  0.6531156560316873  0.2500539122207215
  0.6530557864677249  0.1530502487494374  0.2499692349713882
  0.6530765937871124  0.6530724638196048  0.2500043401234109
  0.0000444025612936  0.3469314680475222  0.2501137782138798
  0.9999968122434879  0.8469233411460380  0.2500045508608437
  0.5000344083972195  0.3469188404153556  0.2500527703590549
  0.4999950720289874  0.8469453620121214  0.2499698271107320
  0.1666019214640855  0.4863761778644999  0.4165981301530834
  0.1667783430925951  0.9866080247307210  0.4165295941128235
  0.6668719127506080  0.4865348433287551  0.4163533896234739
  0.6665503492908442  0.9864048693201326  0.4163153957791792
  0.3197730693090719  0.3333982409778997  0.4165988765288660
  0.3198555861500409  0.8334510640989308  0.4163152821002143
  0.8198295980213857  0.3332219415376514  0.4165293715675773
  0.8196627558403335  0.8331283373902352  0.4163542852039157
\end{lstlisting}
\newpage

\begin{lstlisting}
Cr:AlB4O6N optimised with r2SCAN
1.0000000000000000     
 10.0705041884999993   0.0000000000000000    0.0000000000000000
-5.0352520942999996    8.7213124561999997    0.0000000000000000
 0.0000000000000000    0.0000000000000000   16.4794425963999984
Cr   Al    B     O     N             
1    15    64    96    16
Direct
     0.1666673410598300    0.3333337401808334    0.5301422962752029
     0.1666666847477674    0.3333333185657683    0.0303711652087408
     0.1667389296882288    0.8333694798538709    0.0303644985582935
     0.1666351754291185    0.8333176618366123    0.5303347715962643
     0.6666305351528956    0.3332610895593388    0.0303644974377485
     0.6666823645667748    0.3333648518820798    0.5303347712655019
     0.6666304786624149    0.8333695469609276    0.0303644859458080
     0.6666823012682687    0.8333176323868047    0.5303347732810227
     0.3333983693424528    0.1666016254437748    0.2802136877955977
     0.3334249422583321    0.1665750108150875    0.7805700795904739
     0.3333983868569378    0.6667966914752389    0.2802137176180443
     0.3334249645249739    0.6668498690935354    0.7805697758682444
     0.8332033025956578    0.1666016215623069    0.2802137110031380
     0.8331501042288354    0.1665750830818596    0.7805698996097022
     0.8333333523786940    0.6666666405198854    0.2799578908043552
     0.8333333455835082    0.6666666954060027    0.7805180844410275
     0.4155935238725531    0.0844064856529126    0.1239634536112827
     0.4157893258322286    0.0842106416187676    0.6240685199997814
     0.4156165241968517    0.5844531580821638    0.1239225214088833
     0.4171173729653860    0.5859398742354428    0.6246878019206398
     0.9155468524397238    0.0843834875763553    0.1239225203969173
     0.9140600538973320    0.0828827105538547    0.6246879595406241
     0.9155570671779329    0.5844429371947578    0.1237604706857317
     0.9154700095364112    0.5845298955916158    0.6241205565986634
     0.4156165253751743    0.3311634653944059    0.1239225366041410
     0.4171173515465644    0.3311774830520243    0.6246878889467736
     0.4155934677175737    0.8311871247989153    0.1239634839969312
     0.4157892703762214    0.8315787435536445    0.6240685241078804
     0.9155469232867446    0.3311634517023506    0.1239225630797449
     0.9140598918948526    0.3311776434439090    0.6246877172487882
     0.9155569070785244    0.8311139792278476    0.1237604337146134
     0.9154701478816913    0.8309403793714388    0.6241206733418899
     0.1688365431098346    0.0843834826791829    0.1239225342701825
     0.1688224554224438    0.0828825414287453    0.6246879420148578
     0.1688365643625728    0.5844530929905934    0.1239225631784234
     0.1688223188387643    0.5859398760132063    0.6246877154797271
     0.6688128849384809    0.0844065440745336    0.1239634832209130
     0.6684211867514351    0.0842107770921944    0.6240685282816949
     0.6688860240745077    0.5844430984474649    0.1237604442853288
     0.6690596241982405    0.5845298161750035    0.6241206261674664
     0.0840441344871088    0.4159559558941883    0.3728527787145433
     0.0844011451490488    0.4155988228024467    0.8740144814977350
     0.0842398192990881    0.9152026925088054    0.3736996477620365
     0.0844235038549241    0.9155288996690679    0.8741159209266050
     0.5847973182680973    0.4157601715955195    0.3736996469356704
     0.5844710916393806    0.4155764984801655    0.8741158825676177
     0.5844214339355311    0.9155785669269738    0.3740229494377469
     0.5844116308953258    0.9155884196977718    0.8740790172720168
     0.0840441351859090    0.1680880499233983    0.3728527668783285
     0.0844011481453801    0.1688022058107601    0.8740144978575511
     0.0842398668461866    0.6690370476108991    0.3736996784188373
     0.0844235009884701    0.6688945207285230    0.8741159036211813
     0.5844214372804829    0.1688428099611326    0.3740229707747015
     0.5844115914328010    0.1688231179882842    0.8740790667606906
     0.5847972803135919    0.6690370314857489    0.3736996746716802
     0.5844711440771242    0.6688945493912473    0.8741159378382533
     0.3319121005725919    0.4159558836405933    0.3728526377931285
     0.3311977820373017    0.4155988211143763    0.8740145049768877
     0.3311571965847415    0.9155785691774434    0.3740229694868973
     0.3311768823085855    0.9155883864450427    0.8740790627532461
     0.8309629780273881    0.4157601520871151    0.3736996801691013
     0.8311054928799918    0.4155765348586256    0.8741159066823497
     0.8309629622550579    0.9152026968020051    0.3736996743903192
     0.8311054753985457    0.9155289091118002    0.8741159291594705
     0.9989137637254171    0.9978273408248552    0.4949041534924635
     0.0000082066744196    0.0000163714889110    0.9951297370822202
     0.9989135620394972    0.5010863218785599    0.4949041515782716
     0.0000082388514138    0.4999917802452352    0.9951297510192041
     0.4999999817808374    0.0000000287451924    0.4951344942501186
     0.5000000162194012    0.9999999936642222    0.9951734858496855
     0.5021727847844119    0.5010861895929877    0.4949040339736648
     0.4999836476544806    0.4999917971692997    0.9951297384148020
     0.9997522791256271    0.9995046242846087    0.2448155564857104
     0.9995685597739782    0.9991371677916729    0.7455131638424244
     0.9997523381521202    0.5002476816935443    0.2448155975170696
     0.9995685129256552    0.5004314620385633    0.7455126888684925
     0.4999999710802552    0.0000000324231772    0.2451499843017869
     0.4999999669807830    0.0000000620341053    0.7452636250132563
     0.5004954028043999    0.5002477355573746    0.2448155412361441
     0.5008628211238526    0.5004313879392108    0.7455129793921946
     0.2542399448824131    0.2457600567780072    0.0969182640326940
     0.2563909299545297    0.2436089372683518    0.5990351202392211
     0.2542771437825283    0.7458173668281930    0.0969191260739343
     0.2542429518614674    0.7467777280753709    0.5971723337453113
     0.7541826423324969    0.2457228679923549    0.0969191280893527
     0.7532221262279660    0.2457570828188356    0.5971723519696076
     0.7542235810837969    0.7457764195201201    0.0967745352169906
     0.7542651375183682    0.7457348457736473    0.5970382896583150
     0.2542772694793480    0.0084599307074912    0.0969190634712598
     0.2542430479714923    0.0074654281086717    0.5971722975537046
     0.2542399955939340    0.5084799568864551    0.0969182641939564
     0.2563909462861983    0.5127820356928252    0.5990354894509301
     0.7541825711136684    0.0084600299802033    0.0969190490548856
     0.7532222750759240    0.0074655634222750    0.5971722775844217
     0.7542237287474967    0.5084475097449616    0.0967744830283679
     0.7542651747283347    0.5085305800785734    0.5970381782238439
     0.4915400724312375    0.2457227374864697    0.0969190633445062
     0.4925345073500178    0.2457569277436955    0.5971722679120820
     0.4915399752286808    0.7458174361002416    0.0969190438103209
     0.4925344363296951    0.7467776022191412    0.5971723010588059
     0.9915200570869666    0.2457600144023906    0.0969182598952193
     0.9872177714926453    0.2436091916430468    0.5990354147706776
     0.9915524987014497    0.7457762745337836    0.0967744850360302
     0.9914694094783768    0.7457348472857682    0.5970382013077380
     0.2456874536767412    0.2543124337557588    0.3462347230920969
     0.2457364604498611    0.2542635070214833    0.8469188961530790
     0.2457085156271042    0.7542286855616347    0.3469354345890173
     0.2457752642859164    0.7543274238810438    0.8470518488270168
     0.7457713198338967    0.2542914948908425    0.3469354265580727
     0.7456725622996681    0.2542247461511362    0.8470518759791865
     0.7457470354893561    0.7542529381627330    0.3463703520772831
     0.7457231701117469    0.7542768581558362    0.8469898686558526
     0.2456874335179760    0.4913747962456461    0.3462347789301752
     0.2457364044016264    0.4914726699440298    0.8469188468563941
     0.2457084215875567    0.9914796020114539    0.3469353621288355
     0.2457751409226709    0.9914475894088439    0.8470517612633963
     0.7457470040101082    0.4914938606227511    0.3463702731203132
     0.7457230930190922    0.4914461051570688    0.8469898021789145
     0.7457712501276907    0.9914796005095317    0.3469353646642583
     0.7456725696092183    0.9914476299749495    0.8470517285494750
     0.0086252173724326    0.2543125771648287    0.3462347528463187
     0.0085272935658447    0.2542635598989135    0.8469188406166057
     0.0085061488167775    0.7542530073282185    0.3463702732074159
     0.0085539049369607    0.7542769141109907    0.8469898226843758
     0.5085203843983079    0.2542915750967073    0.3469353685202999
     0.5085523905134793    0.2542248479143889    0.8470517470986234
     0.5085203890030420    0.7542287415002414    0.3469353745065585
     0.5085523947076031    0.7543274911733854    0.8470517802759717
     0.4209880069121083    0.0790120061842921    0.2135701582226416
     0.4210033848543625    0.0789966318909165    0.7136735581017334
     0.4211963339771935    0.5791689739408579    0.2134497050389002
     0.4224740421520886    0.5800484947673509    0.7141273927610593
     0.9208310256246307    0.0788036672071628    0.2134497029167517
     0.9199514913846416    0.0775259415531938    0.7141280464451881
     0.9208590750336769    0.5791409277000493    0.2132651142291492
     0.9207150326635349    0.5792850219140366    0.7137189946181662
     0.4211963083371992    0.3420274511612165    0.2134496996328069
     0.4224739072708100    0.3424256628690519    0.7141280530706879
     0.4209879429265034    0.8419760922805828    0.2135701713416409
     0.4210034087051194    0.8420069675544875    0.7136735282487172
     0.9208311951759726    0.3420274379222201    0.2134497437310749
     0.9199516177292577    0.3424257031791430    0.7141271375071153
     0.9208590196372961    0.8417181856657000    0.2132650959683662
     0.9207151034237261    0.8414303298121570    0.7137191472999682
     0.1579725398317632    0.0788036837515351    0.2134496976249308
     0.1575743186905731    0.0775259914639622    0.7141280694952028
     0.1579725700882572    0.5791688002759330    0.2134497437141095
     0.1575743311104641    0.5800482919990371    0.7141271167180621
     0.6580239163925929    0.0790120584384893    0.2135701723858708
     0.6579930610023483    0.0789965760116354    0.7136735510253895
     0.6582818241332103    0.5791409737715960    0.2132650965466082
     0.6585697259501287    0.5792847826044456    0.7137190889996871
     0.0768531727541377    0.4231470056928129    0.4616361674821129
     0.0790512897764368    0.4209486909600054    0.9635967282390694
     0.0782480263621610    0.9193816533874158    0.4634141658998220
     0.0790572629141292    0.9209420163101855    0.9636545598827374
     0.5806183735636948    0.4217519419438773    0.4634141754752347
     0.5790579892629112    0.4209427375798609    0.9636545490764136
     0.5790308118532236    0.9209691660707502    0.4635919043653238
     0.5790238144087320    0.9209762416997492    0.9636491512716000
     0.0768532362259533    0.1537061473735502    0.4616361066059582
     0.0790512843923733    0.1581024727894212    0.9635967330252387
     0.0782481395130048    0.6588663284799362    0.4634141963783850
     0.0790572183783297    0.6581151212170308    0.9636545725272316
     0.5790307736035470    0.1580616582507506    0.4635919126724916
     0.5790237712296272    0.1580475069772825    0.9636491963379190
     0.5806181162516901    0.6588662517146642    0.4634142037070290
     0.5790579816324907    0.6581151670332380    0.9636545809456804
     0.3462943173045824    0.4231465563571851    0.4616355642860031
     0.3418975276690003    0.4209487045850800    0.9635967363280921
     0.3419383509899442    0.9209692220185803    0.4635919114112775
     0.3419524941818383    0.9209762212694304    0.9636491915278346
     0.8411336995539748    0.4217518570295812    0.4634141922930782
     0.8418848894356830    0.4209427955469330    0.9636545732144584
     0.8411336988897279    0.9193817897489508    0.4634141900692315
     0.8418848314970391    0.9209420333899279    0.9636545708891431
     0.0000039545108316    0.0000079856686174    0.0891599605883953
     0.9998982593883478    0.9997962516120297    0.5892982879257607
     0.0000040232855069    0.4999959897652312    0.0891599912708718
     0.9998978285871820    0.5001017696452817    0.5892982128993067
     0.4999999740152633    0.0000000340007541    0.0892553363223358
     0.4999999473983893    0.0000000012621947    0.5892719389664011
     0.4999920181297591    0.4999960466148327    0.0891599617346142
     0.5002036487053158    0.5001016312946204    0.5892982110069080
     0.9996637267985642    0.9993274016802930    0.3386711272151253
     0.9998780607997321    0.9997560903263104    0.8394681121220865
     0.9996637649450902    0.5003362667429720    0.3386711692504066
     0.9998780830352547    0.5001219215812344    0.8394679389059158
     0.5000000106569672    0.9999999924185156    0.3392477981448671
     0.5000000156378364    0.9999999941561893    0.8393301095901036
     0.5006726395625557    0.5003362868143838    0.3386710873108082
     0.5002439001919968    0.5001219166792807    0.8394680510239019
\end{lstlisting}
\newpage

\begin{lstlisting}
Cr:AlB4O6N optimised with HSE06          
1.0000000000000000     
 10.0433139801000006   0.0000000000000000    0.0000000000000000
-5.0216581482000002    8.6977643763000003    0.0000000000000000
 0.0000000000000000    0.0000000000000000   16.4375457764000004
Cr   Al    B     O     N             
1    15    64    96    16
Direct
  0.1666666658537039  0.3333333227150561  0.5301600340549015
  0.1666665864968238  0.3333334118510791  0.0303864135727494
  0.1667762129517314  0.8333880932190922  0.0304017912401946
  0.1666442343861902  0.8333221242886779  0.5303741450294197
  0.6666119063427161  0.3332237808545102  0.0304017925746862
  0.6666779185831189  0.3333557522071544  0.5303741270761719
  0.6666118481503744  0.8333881478246496  0.0304017381312320
  0.6666778558371220  0.8333221247752221  0.5303742081664851
  0.3334147027815462  0.1665852663443204  0.2802130818674939
  0.3334227162180881  0.1665772288814509  0.7806801990459604
  0.3334147740075508  0.6668296442916102  0.2802130968963255
  0.3334225007332563  0.6668450331332423  0.7806801466095905
  0.8331703234775460  0.1665852012778899  0.2802130941376149
  0.8331549051973255  0.1665775192450170  0.7806801544708648
  0.8333333853765836  0.6666665908084610  0.2798659560664376
  0.8333332781463412  0.6666667568597262  0.7806363666715228
  0.4155804642340684  0.0844195240439660  0.1239922273827432
  0.4157637199603670  0.0842361962191944  0.6241723539425621
  0.4155941509746839  0.5844756844911387  0.1239073183108630
  0.4170344654873617  0.5858992004542785  0.6247598917505002
  0.9155243112157052  0.0844058420839531  0.1239073164465267
  0.9141007015377625  0.0829655134110254  0.6247598775387218
  0.9155447678828281  0.5844552348438796  0.1236633845959716
  0.9154583471555924  0.5845416790986633  0.6242021150536843
  0.4155940267623990  0.3311182287097907  0.1239072774797947
  0.4170348369818413  0.3311347882412221  0.6247598397178109
  0.4155804958923923  0.8311607961282164  0.1239921991986392
  0.4157636878417748  0.8315270530062122  0.6241723956233329
  0.9155243437341909  0.3311180902584425  0.1239072861940258
  0.9141005071018711  0.3311349510543309  0.6247598056874750
  0.9155446378267911  0.8310884069126558  0.1236632054368698
  0.9154583369327298  0.8309165578262991  0.6242020819580176
  0.1688817688473847  0.0844059687572383  0.1239072770901473
  0.1688651941824091  0.0829653113262836  0.6247598575195283
  0.1688819042323715  0.5844756470671797  0.1239072876879845
  0.1688650793768218  0.5858995196613108  0.6247597965871918
  0.6688392012931175  0.0844195018812499  0.1239921990301980
  0.6684729180754019  0.0842363275697480  0.6241723888134345
  0.6689115888312500  0.5844553645134738  0.1236632019762993
  0.6690835005080515  0.5845417113565574  0.6242020794333882
  0.0840291910242570  0.4159707881969510  0.3728623607811912
  0.0844184754263537  0.4155815230937492  0.8740689357947389
  0.0842526227145228  0.9151791207008415  0.3736782481414593
  0.0844521310059747  0.9155448409269127  0.8742234657669243
  0.5848208613101988  0.4157473489292016  0.3736782443878042
  0.5844551504335556  0.4155478374543335  0.8742234784862788
  0.5844237956640299  0.9155761704330629  0.3740594362841705
  0.5844162361354108  0.9155837608515540  0.8742025452282007
  0.0840291489688880  0.1680587571350856  0.3728622954290444
  0.0844185995603013  0.1688374802882251  0.8740688389132174
  0.0842525901932163  0.6690738378095133  0.3736782079652201
  0.0844521662243523  0.6689077097972245  0.8742233650579081
  0.5844238024934896  0.1688479560334457  0.3740593795537350
  0.5844162807759830  0.1688327753582897  0.8742025586355808
  0.5848209419575596  0.6690738722587781  0.3736782170057040
  0.5844551409004595  0.6689076167951740  0.8742233712381733
  0.3319412238457886  0.4159708234103547  0.3728622976363667
  0.3311625192605518  0.4155813816793170  0.8740688400565091
  0.3311520183333627  0.9155761825220239  0.3740593789214586
  0.3311672013671654  0.9155837011392904  0.8742025511451104
  0.8309261469414508  0.4157473887560101  0.3736782104581522
  0.8310922933290712  0.4155478419599774  0.8742233645377908
  0.8309260985635092  0.9151790384734042  0.3736782131535961
  0.8310923687262317  0.9155448709180618  0.8742233769067269
  0.9989092190632221  0.9978186517168766  0.4946671836960306
  0.0000374490626527  0.0000751657047644  0.9949420740092023
  0.9989092310709111  0.5010907879403703  0.4946671837226120
  0.0000373088945764  0.4999626993290960  0.9949420821253057
  0.4999999344024815  0.0000000418002131  0.4949722870752566
  0.4999999227087528  0.0000000684303103  0.9950458225287022
  0.5021813585307413  0.5010907524986550  0.4946671774309337
  0.4999248280993385  0.4999625487913733  0.9949420767938477
  0.9997533374971894  0.9995065196530533  0.2445315813129127
  0.9995835652072813  0.9991667171259522  0.7453680165164087
  0.9997533867001636  0.5002465890323649  0.2445316528369332
  0.9995833813694404  0.5004166481398897  0.7453679640258457
  0.4999999978567615  0.9999999820623273  0.2449485585340767
  0.5000000902503032  0.9999998782906729  0.7451449800248682
  0.5004934627668902  0.5002466443278237  0.2445315788043203
  0.5008333780520360  0.5004163578214147  0.7453680206773683
  0.2543739382473689  0.2456260632089737  0.0970065996411265
  0.2563372252654261  0.2436629058859552  0.5991377344328370
  0.2544166596124953  0.7456912840644776  0.0970342237076736
  0.2543872624375538  0.7465914738996204  0.5973496665076041
  0.7543087212695596  0.2455833340287512  0.0970342232423533
  0.7534085539442330  0.2456127549761362  0.5973496801031928
  0.7543723726332345  0.7456276248376525  0.0967470613945522
  0.7543811835402110  0.7456188586179380  0.5972265845665987
  0.2544166070568465  0.0087250908831038  0.0970342494509211
  0.2543873404140626  0.0077955950952600  0.5973495937471611
  0.2543739829129592  0.5087477329883114  0.0970066133863412
  0.2563371123358849  0.5126741650813145  0.5991373999526175
  0.7543086221506812  0.0087251092308449  0.0970342534174833
  0.7534086506587698  0.0077955759145780  0.5973497849665961
  0.7543726438358505  0.5087440598721571  0.0967470773480130
  0.7543811456026432  0.5087621494039993  0.5972266728215914
  0.4912749035653121  0.2455833832680412  0.0970342494842740
  0.4922043886285081  0.2456125347638647  0.5973495233365895
  0.4912748798046920  0.7456913673216690  0.0970342572034255
  0.4922044185322036  0.7465912880808574  0.5973498250812739
  0.9912522691262211  0.2456260130540429  0.0970066049946681
  0.9873258621403380  0.2436629029032247  0.5991374301411412
  0.9912559739884657  0.7456273770951753  0.0967470576984937
  0.9912378590946531  0.7456188754572608  0.5972266672874511
  0.2456016997634549  0.2543982788616574  0.3463715496453474
  0.2456200796520775  0.2543799323329949  0.8470905722626085
  0.2456399949084513  0.7543556124642308  0.3470813125614640
  0.2456932593925387  0.7544346686359802  0.8473207997909995
  0.7456443589274500  0.2543599746868779  0.3470813085866666
  0.7455653147850398  0.2543067451164518  0.8473208067887015
  0.7456142079789103  0.7543857703097601  0.3463859656702653
  0.7456059129408601  0.7543940751978724  0.8472585433863244
  0.2456017370479842  0.4912038573735344  0.3463715996255630
  0.2456199986078573  0.4912404830442014  0.8470908013837146
  0.2456400248244961  0.9912848014618518  0.3470813482005184
  0.2456933248962869  0.9912588845545045  0.8473209335616687
  0.7456142026997696  0.4912287954851209  0.3463861146586495
  0.7456059068511038  0.4912122430627477  0.8472585849732823
  0.7456443798782502  0.9912847874004669  0.3470813912827992
  0.7455652980659764  0.9912589551888971  0.8473207996804391
  0.0087961106240968  0.2543982323334006  0.3463715970024595
  0.0087595064088291  0.2543800137827859  0.8470908000319142
  0.0087711610232333  0.7543857628984725  0.3463861100445413
  0.0087877291665990  0.7543940570959577  0.8472586004740776
  0.5087151813033799  0.2543599343273328  0.3470813384791143
  0.5087410720690571  0.2543066292952219  0.8473209587200898
  0.5087151847535765  0.7543555817317014  0.3470813978030094
  0.5087410515772604  0.7544346651705780  0.8473207814122006
  0.4211228564505944  0.0788771198793228  0.2134315677700016
  0.4211288012451888  0.0788710678509119  0.7136309566616674
  0.4213312380357763  0.5790592051124577  0.2132653019869366
  0.4225277401624439  0.5798511893731657  0.7140605753399072
  0.9209407788127422  0.0786687477622863  0.2132653036030945
  0.9201484884170839  0.0774721130325986  0.7140605649021339
  0.9209967906249190  0.5790032032372494  0.2129906196870408
  0.9208612704121961  0.5791387482471464  0.7136548592764882
  0.4213311337606811  0.3422718117011385  0.2132653008526404
  0.4225294507678967  0.3426758577755393  0.7140605770817601
  0.4211229183999237  0.8422456352563472  0.2134315659507635
  0.4211288735153360  0.8422572130515817  0.7136310007908904
  0.9209409586922419  0.3422717353184126  0.2132652762137965
  0.9201477583958280  0.3426761339022732  0.7140605475145136
  0.9209970781169474  0.8419929800207981  0.2129904555860946
  0.9208611403175340  0.8417224605247142  0.7136548092422785
  0.1577281768859251  0.0786688576384407  0.2132652955381360
  0.1573240987913138  0.0774709703440806  0.7140605884761015
  0.1577282420212214  0.5790590191904741  0.2132652662059868
  0.1573238971854920  0.5798522570684383  0.7140605410053809
  0.6577543324612449  0.0788770724676340  0.2134315568382448
  0.6577427286601818  0.0788711602700900  0.7136310031450890
  0.6580069906256938  0.5790028986099856  0.2129904572048744
  0.6582776305345703  0.5791389846071411  0.7136548202664486
  0.0768012662847539  0.4231987149989465  0.4615637870681653
  0.0789274996357392  0.4210724987435697  0.9634762364948131
  0.0780679738726207  0.9194807848670479  0.4632461116522322
  0.0789569601368498  0.9211513734800008  0.9635859959370165
  0.5805192073855636  0.4219320000515623  0.4632460971855537
  0.5788486254840777  0.4210430290321625  0.9635859961959952
  0.5788857132975238  0.9211142359862521  0.4634835708888119
  0.5788720671441254  0.9211279027543711  0.9636092169865265
  0.0768010531546608  0.1536027533650994  0.4615637036683395
  0.0789277458325728  0.1578558250632440  0.9634761666272098
  0.0780676641126163  0.6585874623145926  0.4632461080919867
  0.0789568927838147  0.6578059858242398  0.9635859511757658
  0.5788857544980601  0.1577717102115912  0.4634835236575512
  0.5788720971287376  0.1577444784541839  0.9636092584897753
  0.5805195375052818  0.6585874777534215  0.4632461135740584
  0.5788484975668879  0.6578059071637838  0.9635859054789933
  0.3463972404168629  0.4231989222108723  0.4615637119353195
  0.3421441928191626  0.4210722682206693  0.9634761619225642
  0.3422282562571652  0.9211142471117597  0.4634835321365429
  0.3422555141094890  0.9211279139625645  0.9636092525297357
  0.8414125364293952  0.4219323256912304  0.4632461014549492
  0.8421940164841004  0.4210431040428801  0.9635859459285783
  0.8414124947342145  0.9194804583156753  0.4632461118571456
  0.8421940823861505  0.9211514944703580  0.9635859221915055
  0.0000269699328825  0.0000536575701204  0.0890579046140232
  0.9998817834659803  0.9997632987665028  0.5892617072351527
  0.0000270886699312  0.4999729048230392  0.0890579133295049
  0.9998816473477063  0.5001183392601831  0.5892617018664623
  0.4999999984065511  0.9999999587043504  0.0892294906494726
  0.5000001037410229  0.9999998771203877  0.5892789029083332
  0.4999463398091777  0.4999730197832761  0.0890579006505803
  0.5002367502504654  0.5001181937508790  0.5892617031990142
  0.9996592939708506  0.9993188986546571  0.3385434569192967
  0.9998908773142716  0.9997819488980184  0.8394588050405503
  0.9996593452312865  0.5003406288979306  0.3385434771226485
  0.9998907871327987  0.5001092101321802  0.8394588299911945
  0.4999998844109612  0.0000000660696458  0.3391756806963642
  0.4999999316682846  0.0000000556472699  0.8393526629177188
  0.5006810787431348  0.5003406763472640  0.3385434656496003
  0.5002180621088925  0.5001090804167987  0.8394588093330739
\end{lstlisting}
\newpage

\lstinputlisting{structures-for-SI/BeAl2O4_Cs/CONTCAR_BeAl2O4_R2SCAN}
\newpage

\lstinputlisting{structures-for-SI/BeAl2O4_Cs/CONTCAR_BeAl2O4_HSE}
\newpage

\lstinputlisting{structures-for-SI/BeAl2O4_Ci/CONTCAR_BeAl2O4_R2SCAN}
\newpage

\lstinputlisting{structures-for-SI/BeAl2O4_Ci/CONTCAR_BeAl2O4_HSE}
\newpage